\documentclass[aps,prb,preprint,groupedaddress,showpacs]{revtex4-1}

\usepackage{graphicx}
\usepackage{bm}

\begin{document}


\title{Altermagnetism and magnetic groups with pseudoscalar
       electron spin}



\author{Ilja Turek}
\email[]{turek@ipm.cz}
\affiliation{Institute of Physics of Materials,
Czech Academy of Sciences,
\v{Z}i\v{z}kova 22, CZ-616 62 Brno, Czech Republic}


\date{\today}

\begin{abstract}
We revise existing group-theoretical approaches for a treatment
of nonrelativistic collinear magnetic systems with perfect
translation invariance.
We show that full symmetry groups of these systems, which contain
elements with independent rotations in the spin and configuration
spaces (spin groups), can be replaced by magnetic groups consisting
of elements with rotations acting only on position vectors.
This reduction follows from modified transformation properties of
electron spin, which in the considered systems becomes effectively
a pseudoscalar quantity remaining unchanged upon spatial operations
but changing its sign due to an operation of antisymmetry.
We introduce a unitary representation of the relevant magnetic
point groups and use it for a classification of collinear magnets
from the viewpoint of antiferromagnetism-induced spin splitting of
electron bands near the center of Brillouin zone.
We prove that the recently revealed different altermagnetic classes
correspond in a unique way to all nontrivial magnetic Laue classes,
i.e., to the Laue groups containing the operation of antisymmetry
only in combination with a spatial rotation.
Four of these Laue classes are found compatible with a nonzero spin
conductivity.
Subsequent inspection of a simple model allows us to address briefly
the physical mechanisms responsible for the spin splitting in real
systems.
\end{abstract}


\maketitle


\section{Introduction\label{s_intr}}

The most important characteristics of a solid from a viewpoint of 
magnetic properties is certainly its magnetic structure.
A standard classification of various magnetic orders is based on the
mutual arrangement of local magnetic moments and their orientation
with respect to the atomic lattice \cite{r_1982_cmh}.
This approach covers both traditional spin structures (ferromagnets,
spin glasses, etc.) and more exotic orders, such as, e.g., magnetic
skyrmions \cite{r_2016_sm, r_2017_jhh}.
During the last years, the close relation of magnetism and
spintronics gave rise to a complementary approach to the varieties
of magnetic solids, which is based on their electronic structure.
This change of the focus from the real space (local magnetic moments)
to the reciprocal space (electronic spectra) has partly been
motivated by new phenomena related to topological aspects of electron
states \cite{r_2015_orv, r_2018_amv, r_2020_meb} or by a
momentum-dependent spin splitting of electron bands in collinear
antiferromagnets \cite{r_2013_gy, r_2016_non, r_2019_ahl,
r_2019_nhk, r_2019_hyk, r_2021_mkj}.
The latter phenomenon, proposed theoretically by Pekar and Rashba
in 1964 \cite{r_1964_pr}, has recently attracted considerable
attention especially due to the fact that the strength of this
splitting can be sizable also in systems of light elements
\cite{r_2020_ywl, r_2021_ywl_m, r_2021_hsv, r_2021_ssj}.
This contrasts the usual splitting due to spin-orbit interaction,
which is strong mainly in systems containing heavy elements.
The nonrelativistic origin of the antiferromagnetism-induced spin
splitting, a large number of systems exhibiting this property, and
its potential importance for further development of spintronics have
lead to a special term for this type of magnetic order, namely to
altermagnetism, as introduced by L. \v{S}mejkal et al.
\cite{r_2021_ssj, r_2022_ssj}.

In the field of solid-state magnetism, group theory proved its
usefulness several decades ago.
Its standard tools include magnetic groups \cite{r_1965_og,
r_1966_rrb, r_2010_bc} which represent an extension of
crystallographic groups by considering time reversal as
an additional symmetry operation;
the time reversal is a special case of an operation of antisymmetry
or antiidentity contained in some elements of the magnetic groups
\cite{r_1951_avs, r_1964_sb}.
The space-time symmetry in magnetic crystals has well-known
consequences for shape restrictions of various vector or tensor
quantities appearing as equilibrium properties 
\cite{r_1963_rrb, r_1966_rrb} or linear response (transport)
coefficients \cite{r_1966_whk, r_2015_skw, r_2016_wcs, r_2017_zgm}.
This involves, e.g., identification of magnetic point groups
compatible with a net nonzero magnetic moment \cite{r_1965_og}
or modification of the Onsager reciprocity relations for solids
characterized by certain magnetic point groups \cite{r_1966_whk}.
These topics have been worked out to many details, see 
Ref.~\onlinecite{r_2019_gee} and references therein.
Moreover, a scheme for labelling electron eigenvalues in magnetic
crystals, based on irreducible representations of magnetic point
and space groups, is available as well \cite{r_2010_bc}.
This scheme has recently been extended and used in systematic search
for new topological phases of magnetic materials \cite{r_2018_wpv,
r_2020_xes, r_2021_ews}.
The irreducible representations are also indispensable for an
advanced analysis of complex magnetic structures \cite{r_1968_efb,
r_2015_pgt}.

From a viewpoint of electronic structures, treated within effective
one-electron Pauli or Dirac equations, elements of the magnetic
groups act simultaneously on internal degrees of freedom of electron
(spin) and on the electron position vector.
For specific problems, spin groups as an extension of the magnetic
groups were introduced \cite{r_1974_lo, r_1977_dbl}.
Elements of the spin groups are featured by independent rotations
in the spin and configuration spaces.
The spin groups are relevant, e.g., for systems without spin-orbit
interaction; a very recent application of the spin groups deals with
the spin splitting of electron states in collinear antiferromagnets
\cite{r_2021_ssj}.
Undoubtedly, the spin groups comprise all symmetry elements of
nonrelativistic collinear magnets and their use is thus fully
justified.
Nevertheless, one should mention that this extension of theoretical
formalism is accompanied by a substantial increase in the number of
all possible groups: there are 32 crystallographic point
groups, which lead to 122 magnetic point groups \cite{r_1966_whk}
and to 598 nontrivial spin point groups \cite{r_1977_dbl}.
Moreover, inclusion of the translational invariance of crystals
leads to a further extension of the group formalism by considering
the space groups in addition to their point counterparts; this route
has recently been followed with magnetic groups in
Refs.~\onlinecite{r_2020_ywl, r_2021_ywl_m} and with spin groups
in Ref.~\onlinecite{r_2022_llh}.

The more sophisticated formalism of the spin groups as compared with
that of the magnetic groups contradicts obviously the simpler
theoretical and numerical electronic-structure techniques for
nonrelativistic collinear magnets as compared with those for general
magnetic crystals.
The main purpose of this paper is to reconsider the group-theoretical
framework for the electronic structure of nonrelativistic
collinear magnets from the viewpoint of magnetic groups.
We suggest that an alternative treatment of these systems can be
formulated by replacing the vector spin operator by a pseudoscalar
spin quantity, which leads to magnetic groups modified as compared
to those with the standard vector spin.
Such an approach has been mentioned implicitly in the literature
\cite{r_2020_ywl, r_2021_hsv}, but its systematic description is not
available.
In this work, we derive a general unitary infinite-dimensional
representation of the modified magnetic point groups which does not
rely on any particular model of the electronic structure.
We apply the developed formalism to investigation of the spin
splitting of electronic states near the center of the Brillouin
zone (BZ) of nonrelativistic crystalline collinear magnets.
We also study a connection between the spin splitting and spin
conductivity, which has recently lead to a prediction of efficient
spin-current generation \cite{r_2021_hsv} and of giant tunneling
magnetoresistance \cite{r_2021_szl, r_2022_shg}.
Moreover, the obtained results allowed us to address briefly
the physical mechanisms responsible for appearance of this spin
splitting in real materials.

\section{Formalism\label{s_form}}

\subsection{Pseudoscalar electron spin \label{ss_pes}}

Electrons are charged fermions of spin 1/2.
In one-particle approximations for many-electron spin-polarized
systems, the Pauli exclusion principle and the Coulomb interaction
between the electrons give rise to a vector exchange field coupled
to the vector spin operator in the Zeeman term of an effective
one-electron Hamiltonian.
The additional spin-orbit interaction and/or the noncollinear
spin structure (and, consequently, the noncollinear exchange
field) lead to coupled equations for the electron wavefunctions
in the two spin channels (spin-up and spin-down channels) of the
Pauli equation as a nonrelativistic limit of the Dirac equation
\cite{r_1998_ps}.
Transformations of the wavefunctions, comprised in the magnetic
space and point groups, take thus the vector nature of the electron
spin, of the exchange field, and of the electron position vector
fully into account.
Transformation properties of the spin and of the exchange field
are the same as those of the angular orbital momentum 
$\mathbf{r} \times \mathbf{p}$, where $\mathbf{r}$ is the position
vector and $\mathbf{p}$ is the electron momentum.

The situation simplifies substantially for systems with neglected
spin-orbit interaction and with collinear exchange fields
leading thus to collinear spin structures.
The wavefunction amplitudes are $\langle \mathbf{r} s | \psi \rangle
 = \psi_s(\mathbf{r})$, where $s$ denotes the spin index
($s = 1$ for spin-up channel, $s = -1$ for spin-down channel).
The Hamiltonian can be written (in atomic units with $\hbar = 1$
and with the electron mass $m = 1/2$) as
\begin{equation}
H(\mathbf{r}) = - \Delta + V(\mathbf{r}) ,
\label{eq_hrdef}
\end{equation}
where the kinetic energy term is spin-independent, whereas the
local potential $V(\mathbf{r})$ is spin-dependent but diagonal in
the spin index: $\langle s | V(\mathbf{r}) | s' \rangle =
\delta_{ss'} V_s(\mathbf{r})$.
This leads to two eigenvalue problems with eigenvalues $E_s$, 
\begin{equation}
- \Delta \psi_s(\mathbf{r}) + V_s(\mathbf{r}) \psi_s(\mathbf{r})
= E_s \psi_s(\mathbf{r}) , 
\label{eq_hevp}
\end{equation}
to be solved separately in each spin channel ($s = \pm 1$).
If we introduce a spin operator $\sigma$ such that 
$\langle s | \sigma | s' \rangle = s \delta_{ss'}$, a spin-averaged
potential
$\bar{V}(\mathbf{r}) = [ V_+(\mathbf{r}) + V_-(\mathbf{r}) ] / 2$,
and an exchange field
$B(\mathbf{r}) = [ V_+(\mathbf{r}) - V_-(\mathbf{r}) ] / 2$,
the Hamiltonian (\ref{eq_hrdef}) can be rewritten as
\begin{equation}
H(\mathbf{r}) = - \Delta + \bar{V}(\mathbf{r})
 + B(\mathbf{r}) \sigma .
\label{eq_hbsigma}
\end{equation}
The direction of the spin quantization axis is irrelevant, the
Hamiltonian $H(\mathbf{r})$ describes motion in two uncoupled spin
channels with local potentials $V_s(\mathbf{r})$, $s = \pm 1$,
and the defined spin $\sigma$ and exchange field $B(\mathbf{r})$ 
can be treated as scalar quantities.

In magnetic crystals, the Hamiltonian $H(\mathbf{r})$ is
translationally invariant, so that $V(\mathbf{r}) =
V(\mathbf{r} + \mathbf{T})$ for all $\mathbf{r}$ and for all
primitive translation vectors $\mathbf{T}$ (vectors of the Bravais
lattice), which implies the same condition for $V_s(\mathbf{r})$,
$s = \pm 1$, $\bar{V}(\mathbf{r})$, and $B(\mathbf{r})$.
Let us consider further symmetry elements of the system.
In ferromagnets, the two potentials $V_s(\mathbf{r})$ are mutually
different, since $V_+(\mathbf{r})$ is on average more (or less)
attractive than $V_-(\mathbf{r})$.
The system is thus invariant only with respect to ordinary rotations
(combined optionally with nonprimitive translations)
that belong to the crystallographic point group.
These rotations will be denoted by a symbol $\alpha$, which is a
real $3\times 3$ orthogonal matrix, $\alpha \equiv
\{ \alpha_{\mu \nu} \}$,
where the subscripts $\mu$ and $\nu$ denote the Cartesian index
($\mu, \nu \in \{ x, y, z \}$); the rotations $\alpha$ can be both
proper and improper (accompanied by space inversion).

In antiferromagnets, both spin channels are mutually equivalent,
which points to a presence of more general symmetry elements as
compared to ferromagnets.
These elements of the system point group will be denoted as
$(\alpha , \eta)$, where the extra parameter $\eta$ acquires two
values, namely, $\eta = 1$ for symmetry elements not changing the
spin channels, while $\eta = -1$ for symmetry elements with
mutual interchange of both spin channels.
All these elements form the magnetic point group
$\mathcal{P}_\mathrm{M}$ of the system with a group multiplication
rule 
\begin{equation}
(\alpha_1 , \eta_1) (\alpha_2 , \eta_2) = 
(\alpha_1 \alpha_2 , \eta_1 \eta_2) .
\label{eq_mgmr}
\end{equation}
Strictly defined, $(\alpha , \eta) \in \mathcal{P}_\mathrm{M}$
means that a translation vector $\mathbf{t}$ (either null or
nonprimitive) exists such, that
\begin{equation}
V_s(\mathbf{r}) = V_{\eta s}(\alpha \mathbf{r} + \mathbf{t})
\label{eq_mpgdef}
\end{equation}
holds for all $\mathbf{r}$ and for both values of $s$ ($s = \pm 1$).
Hence the group elements $(\alpha , 1)$ correspond to usual
rotations, whereas the group elements $(\alpha , -1)$ correspond to
rotations combined with the spin-channel interchange, which plays a
role of the operation of antisymmetry of the magnetic group
\cite{r_2010_bc}.
Note that the spin-channel interchange does not change only the sign
of the spin channel ($s \to -s$), but it changes the sign of the
exchange field as well [$B(\mathbf{r}) \to -B(\mathbf{r})$].
The electron spin and the exchange field behave thus like
pseudoscalar quantities changing their signs due to the operation of
antisymmetry.
In antiferromagnets, the regions of positive and negative values of
the exchange field $B(\mathbf{r})$ represent an analogy to white and
black regions, respectively, of two-color figures with a symmetry
group extended by inclusion of an operation of antisymmetry
(interchange of colors), as introduced by A. Shubnikov
\cite{r_1951_avs, r_1964_sb}.
However, the group $\mathcal{P}_\mathrm{M}$ defined by
Eq.~(\ref{eq_mpgdef}) reflects the symmetry of both local potentials
$V_s(\mathbf{r})$ ($s = \pm 1$), not only of their difference (the
exchange field), in full compatibility with the density-functional
theory of nonrelativistic collinear magnets \cite{r_1976_gl}.
This means that the presence and positions of nonmagnetic atoms in
the system have to be taken into account in a reliable symmetry
analysis.

The magnetic point groups $\mathcal{P}_\mathrm{M}$ derived from
crystallographic point groups $\mathcal{P}$ can be split
into three categories (a), (b), and (c) \cite{r_1966_whk} or, 
alternatively, into three types I, II, and III \cite{r_2010_bc}
[whereby the categories (a), (b), and (c) correspond to the types
II, I, and III, respectively].
The category (a) comprises all 32 groups $\mathcal{P}$ to which the
operation of antisymmetry is added (so that the pure operation of
antisymmetry is an element of $\mathcal{P}_\mathrm{M}$).
The groups of the category (b) do not involve the operation of
antisymmetry at all (neither as a separate element nor in a
combination with a rotation); all these groups are thus equivalent
to all bare 32 groups $\mathcal{P}$.
The groups $\mathcal{P}_\mathrm{M}$ of the category (c) contain the
operation of antisymmetry only in a combination with a nontrivial
rotation; there are 58 groups in this category.
Each group $\mathcal{P}_\mathrm{M}$ of the category (c) can be
constructed from a parent group $\mathcal{P}$ by taking its subgroup
$\mathcal{S}$ of index two.
All elements $\alpha \in \mathcal{S}$ enter then the group
$\mathcal{P}_\mathrm{M}$ as $(\alpha, 1)$, i.e., without the
operation of antisymmetry, whereas all elements $\alpha \in
\mathcal{P}$ and $\alpha \notin \mathcal{S}$ give rise to elements
containing the operation of antisymmetry, $(\alpha, -1) \in
\mathcal{P}_\mathrm{M}$.
Loosely speaking, the group $\mathcal{S}$ can be identified with
a subgroup of $\mathcal{P}_\mathrm{M}$ containing all elements
of $\mathcal{P}_\mathrm{M}$ without the operation of antisymmetry.
For the magnetic point groups $\mathcal{P}_\mathrm{M}$ defined by 
Eq.~(\ref{eq_mpgdef}), the mentioned three categories are related
unambiguously to basic types of collinear nonrelativistic magnets:
ferromagnets and ferrimagnets possess $\mathcal{P}_\mathrm{M}$ of
category (b), whereas antiferromagnets are featured by
$\mathcal{P}_\mathrm{M}$ of category (a) or (c).
This simple classification contrasts that based on the standard
magnetic groups applied to general magnets (with spin-orbit
coupling and/or with noncollinear orders), where the magnetic point
groups of ferromagnets and ferrimagnets belong to categories (b) and
(c) while those of antiferromagnets belong to categories (a), (b),
and (c).

The magnetic point group $\mathcal{P}_\mathrm{M}$ defined by 
Eq.~(\ref{eq_mpgdef}) can differ from the standard magnetic point
group of the same collinear system.
The latter group reflects the vector nature of the quantities
involved and it depends on the direction of the exchange field and
magnetic moments.
Moreover, the operation of antisymmetry contained in elements of the
standard magnetic groups denotes the time reversal leading to the
sign change of the spin, exchange field, and magnetic moments.
The modification of the magnetic groups owing to the pseudoscalar
nature of the involved quantities can lead to additional spatial
operations contained in the group elements, while the operation of
antisymmetry has to be identified with the spin-channel interchange
according to Eq.~(\ref{eq_mpgdef}).
More details about the relation of both kinds of magnetic point
groups can be found in Ref.~\onlinecite{r_sumampg} and examples of
these groups for selected systems are presented in
Section~\ref{ss_mpgss}.

Let us note that the symmetry operations of the introduced modified
magnetic groups rest on the neglect of all interactions leading to
a coupling of the spin-up and spin-down channels of the one-electron
Hamiltonian. 
In the case of collinear ferromagnets and ferrimagnets, this means
the neglect of spin-orbit interaction and of its well-known
consequences, such as, e.g., the anisotropic magnetostriction often
responsible for reduced symmetry of the lattice in the magnetically
ordered phase as compared to that in the paramagnetic phase 
\cite{r_1995_ava, r_2005_owh}.
This approximate approach resulted in important theoretical concepts
including the half-metallic magnetism \cite{r_1983_gme} or the
symmetry-induced spin filtering in Fe$|$MgO$|$Fe magnetic tunnel
junctions \cite{r_2001_bzs}.
For antiferromagnetis, this approach also neglects a weak
noncollinearity of the magnetic moments in noncentrosymmetric systems
owing to the Dzyaloshinskii-Moriya interaction \cite{r_1958_id,
r_1960_tm}.
The symmetry analysis of nonrelativistic collinear antiferromagnets
has recently been carried out in several theoretical studies using
the spin groups \cite{r_2021_ssj, r_2022_ssj, r_2022_llh}.
These and similar studies are devoted not only to systems of very
light elements, such as MnF$_2$ \cite{r_2020_ywl}, CuF$_2$
\cite{r_2021_ssj}, Mn$_5$Si$_3$ \cite{r_2022_shg}, or NiO
\cite{r_2021_ywl_b, r_2021_ywl_m}, 
but also to systems with heavier elements, such as RuO$_2$
\cite{r_2021_hsv, r_2021_szl}, KRu$_4$O$_8$ \cite{r_2021_ssj},
FeSb$_2$ \cite{r_2021_mkj}, CrSb and MnTe \cite{r_2021_hsv,
r_2021_ssj}, La$_2$CuO$_4$ \cite{r_2021_ssj}, and $A$MnBi$_2$
($A$ = Ca, Sr) \cite{r_2022_llh}.
A comparison of theoretical results of the above approximate
treatment with those of a more accurate description (with spin-orbit
interaction switched on), supported by \emph{ab initio} electronic
structure calculations, enables one to identify the origin of
unusual properties of altermagnetic materials \cite{r_2021_ssj,
r_2022_ssj}.

The magnetic group introduced according to Eq.~(\ref{eq_mpgdef})
contains only symmetry elements for invariance of the pair of 
potentials $V_s(\mathbf{r})$ ($s = \pm 1$).
However, the full group for invariance of the Hamiltonian,
Eq.~(\ref{eq_hrdef}), is inevitably bigger; two additional symmetry
operations have to be considered.
First, it is the spin operator $\sigma$ which commutes obviously with
the Hamiltonian $H(\mathbf{r})$.
This symmetry reflects invariance with respect to arbitrary rotations
in the spin space around the axis parallel to the direction of all
magnetic moments of the collinear magnet. 
Second, the Hamiltonian eigenvalue problem~(\ref{eq_hevp}) is
invariant with respect to complex conjugation of the wave functions:
$\psi_s(\mathbf{r}) \to \psi^\ast_s(\mathbf{r})$.
This symmetry reflects real values of the potentials
$V_s(\mathbf{r}) = V^\ast_s(\mathbf{r})$ and it corresponds to time
reversal for effective particles of spin zero \cite{r_2020_meb,
r_1961_am, r_1959_epw} moving in both decoupled spin channels.
A closer inspection of a relation between the introduced magnetic
groups and the spin groups of the studied systems \cite{r_sumampg}
proves that no further independent symmetry operations exist.
In the following, none of both mentioned additional symmetries
(present in all collinear nonrelativistic magnets) is included in
the magnetic point group $\mathcal{P}_\mathrm{M}$; however, their
possible effect on the results of the performed analysis is taken
properly into account.

\subsection{Hamiltonians and resolvents in reciprocal
            space\label{ss_hrrs}}

In the analysis of spin splitting of the eigenvalues of the
real-space Hamiltonian (\ref{eq_hrdef}), we employ the Bloch
theorem, transform the original $H(\mathbf{r})$ into a
$\mathbf{k}$-dependent Hamiltonian $\tilde{H}(\mathbf{k})$,
where $\mathbf{k}$ denotes a reciprocal-space vector, and focus on
a neighborhood of the center of BZ, i.e., on $\mathbf{k} \to
\mathbf{0}$.
The Hamiltonians $\tilde{H}(\mathbf{k})$ for different $\mathbf{k}$
vectors are defined on different Hilbert spaces.
However, we represent each $\tilde{H}(\mathbf{k})$ by a matrix in an
orthonormal basis $\{ | \mathbf{G} s \rangle \}$, where $\mathbf{G}$
runs over all lattice vectors of the reciprocal lattice and $s$ runs
over both spin channels, $s = \pm 1$.
The basis vectors are chosen as $| \mathbf{G} s \rangle =
| \mathbf{G} \rangle \otimes | s \rangle$, where
$| \mathbf{G} \rangle$ describes a plane wave,
$\langle \mathbf{r} | \mathbf{G} \rangle
\sim \exp[ i ( \mathbf{k} - \mathbf{G} ) \cdot \mathbf{r} ]$,
and where $| s \rangle$ denotes the basis vector in the
two-dimensional spin space.
This plane-wave basis is used in a formulation of the nearly-free
electron model \cite{r_2020_meb, r_2001_js}; however, it leads to
accurate eigenvalues as long as the full infinite basis set
$\{ | \mathbf{G} s \rangle \}$ is employed.
With this matrix representation, all Hamiltonians can be considered
as defined on the same Hilbert space $\mathcal{H}$ (corresponding
to $\mathbf{k} = \mathbf{0}$).
The particular form of $\tilde{H}(\mathbf{k})$ is given in
Appendix \ref{app_hto}.
Its full dependence on $\mathbf{k}$ is confined to a few terms,
\begin{eqnarray}
\tilde{H}(\mathbf{k}) & = & h + U(\mathbf{k}) ,
\nonumber\\
U(\mathbf{k}) & = & \sum_\mu J_\mu k_\mu +
\sum_{\mu_1 \mu_2} L_{\mu_1 \mu_2} k_{\mu_1} k_{\mu_2} .
\label{eq_hktayl}
\end{eqnarray}
The first term $h$ refers to the Hamiltonian for $\mathbf{k} =
\mathbf{0}$ and the operator $U(\mathbf{k})$, consisting of terms
linear and quadratic in $\mathbf{k}$, can be considered for
$\mathbf{k} \to \mathbf{0}$ as a small perturbation added to the
reference Hamiltonian $h$.
The operators $J_\mu$ coincide with components of a velocity
operator and the operators $L_{\mu_1 \mu_2}$ are symmetric in
their indices, $L_{\mu_1 \mu_2} = L_{\mu_2 \mu_1}$.
The latter equal to $L_{\mu_1 \mu_2} = I \delta_{\mu_1 \mu_2}$,
where $I$ is the unit operator in $\mathcal{H}$.

The spin-resolved eigenvalues $E^{(n)}_s(\mathbf{k})$, where
$n$ denotes the band index, depend on the matrix elements of
$\tilde{H}(\mathbf{k})$ in a complicated manner.
Moreover, a thorough analysis of the spin splitting requires a
reliable identification of the spin pairs of eigenvalues, which is
not always straightforward owing to band crossing
\cite{r_2021_ywl_b}.
In order to avoid these problems, we turn to techniques developed
earlier for shapes of various tensor quantities due to the point
group symmetry of the system \cite{r_1963_rrb, r_1966_whk,
r_2015_skw}.
For this purpose, we focus on spin-resolved Bloch spectral functions
$A_s(\mathbf{k},E) = \sum_n \delta(E - E^{(n)}_s(\mathbf{k}))$,
where $E$ denotes an energy variable.
Let us note that the Bloch spectral functions substitute the energy
bands in strongly correlated systems \cite{r_2019_ahl}.
The spin splitting of the system eigenvalues is reflected by
nonzero values of the difference $A_+(\mathbf{k},E) - 
A_-(\mathbf{k},E) = \sum_s s A_s(\mathbf{k},E)$.
The Bloch spectral functions are closely related to the resolvent
$G(\mathbf{k}, E \pm i \varepsilon)$ of the Hamiltonian
$\tilde{H}(\mathbf{k})$, defined for $\varepsilon > 0$ by
\cite{r_1992_ag}
\begin{equation}
G(\mathbf{k}, E \pm i \varepsilon) = 
[ (E \pm i \varepsilon) I - \tilde{H}(\mathbf{k} ]^{-1} .
\label{eq_gkedef}
\end{equation}
This yields explicit relations involving the quantity
$\sum_s s A_s(\mathbf{k},E)$:
\begin{eqnarray}
\mathrm{Tr} [ \sigma G(\mathbf{k}, E \pm i \varepsilon) ] & = &
\int_{-\infty}^{+\infty} \frac{1}{ E \pm i \varepsilon - E' }
\sum_s s A_s( \mathbf{k} , E' ) dE' ,
\nonumber\\
\sum_s s A_s(\mathbf{k},E) & = & - \frac{1}{\pi} \Im \mathrm{Tr}
 [ \sigma G(\mathbf{k}, E + i 0) ] ,
\label{eq_sas}
\end{eqnarray}
where $\Im$ denotes imaginary part and the trace $\mathrm{Tr}$
refers to the Hilbert space $\mathcal{H}$.
In the following, we thus examine the properties of 
$\mathrm{Tr} [ \sigma G(\mathbf{k}, E \pm i \varepsilon) ]$ for
small $\mathbf{k}$ vectors.

Let us denote the resolvent of the reference Hamiltonian $h$ 
as $g(E \pm i \varepsilon)$ and let us employ it in evaluation
of the $\mathbf{k}$-dependent resolvent $G(\mathbf{k} ,
 E \pm i \varepsilon)$.
For brevity, we omit the energy arguments of both resolvents.
The infinite Born series corresponding to Eq.~(\ref{eq_hktayl}),
\begin{equation}
G(\mathbf{k}) = g + \sum_{N \ge 1} [ g U(\mathbf{k}) ]^N g ,
\label{eq_born}
\end{equation}
can be rearranged into the Taylor series
\begin{equation}
G(\mathbf{k}) = g + \sum_{N \ge 1} \sum_{\mu_1 \mu_2 \dots \mu_N}
g W^{(N)}_{\mu_1 \mu_2 \dots \mu_N} g k_{\mu_1} k_{\mu_2} \dots
k_{\mu_N} ,
\label{eq_gktay}
\end{equation}
where the operators $W^{(N)}_{\mu_1 \mu_2 \dots \mu_N}$ are fully
symmetric in the indices $\mu_1, \dots , \mu_N$.
The first four members of the infinite sequence
$W^{(N)}_{\mu_1 \mu_2 \dots \mu_N}$, $N = 1, 2, \dots$, equal to
\begin{eqnarray}
W^{(1)}_\mu & = & J_\mu,
\qquad
W^{(2)}_{\mu_1 \mu_2} = 
L_{\mu_1 \mu_2} + \frac{1}{2} \left(
J_{\mu_1} g J_{\mu_2} + J_{\mu_2} g J_{\mu_1} \right) ,
\nonumber\\
W^{(3)}_{\mu_1 \mu_2 \mu_3} & = & \frac{1}{6} (
J_{\mu_1} g L_{\mu_2 \mu_3} + L_{\mu_1 \mu_2} g J_{\mu_3} +
J_{\mu_1} g J_{\mu_2} g J_{\mu_3} + \dots ) ,
\nonumber\\
W^{(4)}_{\mu_1 \mu_2 \mu_3 \mu_4} & = & \frac{1}{24} (
L_{\mu_1 \mu_2} g L_{\mu_3 \mu_4} +
J_{\mu_1} g J_{\mu_2} g L_{\mu_3 \mu_4} +
J_{\mu_1} g L_{\mu_2 \mu_3} g J_{\mu_4} +
\nonumber\\
 & & {} + L_{\mu_1 \mu_2} g J_{\mu_3} g J_{\mu_4} +
J_{\mu_1} g J_{\mu_2} g J_{\mu_3} g J_{\mu_4} + \dots ) ,
\label{eq_w1234}
\end{eqnarray}
where the dots denote terms obtained from the given ones
by all permutations of the indices $\mu_1, \mu_2, \dots$.
The infinite series (\ref{eq_gktay}) leads to the following
Taylor expansion of the quantity $F(\mathbf{k}) = \mathrm{Tr}
[ \sigma G(\mathbf{k}) ]$:
\begin{eqnarray}
F(\mathbf{k}) & = & \mathrm{Tr} ( \sigma g ) +
\sum_{N \ge 1} \sum_{\mu_1 \mu_2 \dots \mu_N}
T^{(N)}_{\mu_1 \mu_2 \dots \mu_N}
k_{\mu_1} k_{\mu_2} \dots k_{\mu_N} ,
\nonumber\\
T^{(N)}_{\mu_1 \mu_2 \dots \mu_N} & = &
\mathrm{Tr} [ \sigma g W^{(N)}_{\mu_1 \mu_2 \dots \mu_N} g ] ,
\label{eq_tayfk}
\end{eqnarray}
where the tensor components $T^{(N)}_{\mu_1 \mu_2 \dots \mu_N}$
are fully symmetric in their indices.
We will investigate the shape of the tensors $T^{(N)}$ due to
the symmetry of the studied system; nonvanishing components
$T^{(N)}_{\mu_1 \mu_2 \dots \mu_N}$ correspond to spin splitting
of energy bands near the BZ center.

\subsection{Representation of magnetic point
            groups\label{ss_rmpg}}

Since the operators $h$, $J_\mu$, $L_{\mu_1 \mu_2}$, $g$, and 
$W^{(N)}_{\mu_1 \mu_2 \dots \mu_N}$, involved in the expansions
(\ref{eq_hktayl}), (\ref{eq_gktay}), and (\ref{eq_tayfk}), act
in the Hilbert space $\mathcal{H}$ for zero $\mathbf{k}$ vector,
the symmetry analysis can be carried out in terms of the
magnetic point group $\mathcal{P}_\mathrm{M}$ of the system.
For this purpose, one has to construct the corresponding
representation of the group $\mathcal{P}_\mathrm{M}$ by means of
operators $\mathcal{D}(\alpha, \eta)$ acting in the space
$\mathcal{H}$ \cite{r_2010_bc, r_1960_vh, r_1979_ed}.
The spatial parts of the orthonormal basis vectors
$| \mathbf{G} s \rangle$ for $\mathbf{k} = \mathbf{0}$ are given
by $\langle \mathbf{r} | \mathbf{G} \rangle
\sim \exp( - i \mathbf{G} \cdot \mathbf{r} )$ and we define the
unitary operators $\mathcal{D}(\alpha, \eta)$ explicitly by
\begin{equation}
\mathcal{D}(\alpha, \eta) | \mathbf{G} s \rangle =
| \alpha \mathbf{G}, \eta s \rangle
\exp( i \alpha \mathbf{G} \cdot \mathbf{t} ) ,
\label{eq_drepdef}
\end{equation}
where $\mathbf{t}$ denotes the translation vector involved in
the invariance condition (\ref{eq_mpgdef}).
Note that this definition includes naturally the rotation of the
reciprocal lattice vectors ($\mathbf{G} \to \alpha \mathbf{G}$) and
the sign change of the spin index ($s \to \eta s$) due to the
operation of antisymmetry.
The additional phase factor in Eq.~(\ref{eq_drepdef}) is consistent
with a general rule for rotations and translations in a space of
scalar functions of the position vector $\mathbf{r}$ 
\cite{r_2010_bc, r_1979_ed}. 
Alternatively, one can show that Eq.~(\ref{eq_drepdef}) follows
from a simple transformation of all basic kets
$| \mathbf{r}s \rangle$ due to a combined effect of the rotation
$\alpha$, translation $\mathbf{t}$, and spin-channel interchange
$\eta$, which yields $| \mathbf{r}s \rangle \to | \mathbf{r}'s'
\rangle$, where $\mathbf{r}' = \alpha \mathbf{r} + \mathbf{t}$ and
$s' = \eta s$, see also Eq.~(\ref{eq_sm2_deltars}) in
Ref.~\onlinecite{r_sumampg}.
It can be proved that the introduced operators
$\mathcal{D}(\alpha, \eta)$, Eq.~(\ref{eq_drepdef}), possess all
properties of a representation; in particular, the operator
counterpart of the group multiplication rule (\ref{eq_mgmr}),
\begin{equation}
\mathcal{D}(\alpha_1, \eta_1) \mathcal{D}(\alpha_2, \eta_2)
 = \mathcal{D}(\alpha_1 \alpha_2, \eta_1 \eta_2) ,
\label{eq_repbr}
\end{equation}
holds for all elements
$(\alpha_1, \eta_1) \in \mathcal{P}_\mathrm{M}$ and
$(\alpha_2, \eta_2) \in \mathcal{P}_\mathrm{M}$ (for a proof,
see Appendix \ref{app_hto}).

Let us compare briefly the present treatment of rotations and of
the operation of antisymmetry according to Eq.~(\ref{eq_drepdef})
with other group-theoretical approaches.
Elements of the spin groups contain two independent rotations, acting
separately in the spin and configuration spaces \cite{r_1974_lo,
r_2021_ssj, r_2022_llh}, in contrast to the rotations of standard
magnetic groups, acting simultaneously in both spaces
\cite{r_1965_og, r_2010_bc}.
However, the standard magnetic point groups applied to one-particle
Hamiltonians for real electrons with spin 1/2 lead to
double-valued representations \cite{r_1960_vh, r_1979_ed, r_2010_bc}.
Moreover, the operation of antisymmetry is identified with time
reversal and the group elements containing the time reversal are
represented by antiunitary operators, which calls for the use of
co-representations of these magnetic groups \cite{r_2010_bc}. 
The present formalism does not employ any of these extensions of the
group theory.
The structure of the nonrelativistic Hamiltonian for collinear
magnets (Section \ref{ss_pes}) allows one to confine the action of
rotations only to the configuration space, while the operation of
antisymmetry reduces to the interchange of the spin channels,
see Eq.~(\ref{eq_mpgdef}).
As a consequence, the defined representation
$\mathcal{D}(\alpha, \eta)$, Eq.~(\ref{eq_drepdef}), is single-valued
and all elements $(\alpha, \eta)$ of the modified magnetic point
groups $\mathcal{P}_\mathrm{M}$ are represented by unitary operators,
so that no co-representations have to be considered.
These features simplify the formalism substantially. 

The introduced representation (\ref{eq_drepdef}) leads to the
following transformations of the involved operators.
For each element $(\alpha, \eta) \in \mathcal{P}_\mathrm{M}$
and with abbreviation $\mathcal{D}(\alpha, \eta) = D$, we get:
\begin{eqnarray}
D^{-1} h D & = & h ,
\qquad 
D^{-1} g D = g ,
\qquad
D^{-1} \sigma D = \eta \sigma ,
\nonumber\\
D^{-1} J_\mu D & = & \sum_\nu \alpha_{\mu \nu} J_\nu ,
\qquad
D^{-1} L_{\mu_1 \mu_2} D = \sum_{\nu_1 \nu_2}
\alpha_{\mu_1 \nu_1} \alpha_{\mu_2 \nu_2} L_{\nu_1 \nu_2} ,
\nonumber\\
D^{-1} W^{(N)}_{\mu_1 \mu_2 \dots \mu_N} D & = &
\sum_{\nu_1 \nu_2 \dots \nu_N}
\alpha_{\mu_1 \nu_1} \alpha_{\mu_2 \nu_2} \dots \alpha_{\mu_N \nu_N}
W^{(N)}_{\nu_1 \nu_2 \dots \nu_N} .
\label{eq_trpall}
\end{eqnarray}
The proof of these relations is sketched in Appendix \ref{app_hto}
and their physical meaning is obvious:
the Hamiltonian $h$ and the resolvent $g$ are invariant with respect
to action of all group elements, the velocity operators $J_\mu$ are
components of a vector operator, the operators $L_{\mu_1 \mu_2}$ and 
$W^{(N)}_{\mu_1 \mu_2 \dots \mu_N}$ are components of tensor
operators of rank 2 and $N$, respectively, and the spin $\sigma$
changes its sign due to the operation of antisymmetry, but remains
unchanged by pure spatial rotations, in full agreement with its
pseudoscalar nature discussed in Section \ref{ss_pes}.

The time reversal mentioned in the last paragraph of Section
\ref{ss_pes} has to be represented by an antiunitary operator.
We denote it by $\mathcal{T}$ and define it explicitly by
\begin{equation}
\mathcal{T} | \mathbf{G} s \rangle = 
 | \! - \! \mathbf{G}, s \rangle ,
\label{eq_trepdef}
\end{equation}
so that $\mathcal{T}$ changes the sign of the reciprocal lattice
vector $\mathbf{G}$, but leaves the spin index $s$ unchanged.
We have thus $\mathcal{T}^2 = I$ and obtain the following
transformation rules:
\begin{eqnarray}
\mathcal{T} h \mathcal{T} & = & h ,
\qquad 
\mathcal{T} g \mathcal{T} = g^+ , 
\qquad 
\mathcal{T} \sigma \mathcal{T} = \sigma ,
\nonumber\\
\mathcal{T} J_\mu \mathcal{T} & = & - J_\mu ,
\qquad 
\mathcal{T} L_{\mu_1 \mu_2} \mathcal{T} = L_{\mu_1 \mu_2} ,
\nonumber\\
\mathcal{T} W^{(N)}_{\mu_1 \mu_2 \dots \mu_N} \mathcal{T}
 & = & (-1)^N [ W^{(N)}_{\mu_1 \mu_2 \dots \mu_N} ]^+ ,
\label{eq_trttr}
\end{eqnarray}
where $M^+$ denotes the Hermitian conjugate of an operator $M$.
Note especially the unchanged sign of the spin operator $\sigma$,
which reflects the fact that the time reversal treats each separate
spin channel as a subspace for a particle of spin zero.
The sign $(-1)^N$ in the transformation of operators
$W^{(N)}_{\mu_1 \mu_2 \dots \mu_N}$ is due to the velocities $J_\mu$
in their definition (\ref{eq_w1234}).

\subsection{Shape analysis of the studied tensors\label{ss_sast}}

The invariance of the system with respect to the time reversal
(\ref{eq_trttr}) has an obvious consequence for the studied tensors
$T^{(N)}$.
We get from Eq.~(\ref{eq_tayfk}) for $N$ odd:
\begin{eqnarray}
T^{(N)}_{\mu_1 \mu_2 \dots \mu_N} & = & - \mathrm{Tr} \{ 
\mathcal{T} \sigma \mathcal{T} \mathcal{T} g^+ \mathcal{T} 
\mathcal{T} [ W^{(N)}_{\mu_1 \mu_2 \dots \mu_N} ]^+ \mathcal{T} 
\mathcal{T} g^+ \mathcal{T} \} 
\nonumber\\
 & = & - \mathrm{Tr} \{ g W^{(N)}_{\mu_1 \mu_2 \dots \mu_N}
 g \sigma \} = - T^{(N)}_{\mu_1 \mu_2 \dots \mu_N} , 
\label{eq_trtnt}
\end{eqnarray}
where we used the rule $\mathrm{Tr} ( \mathcal{T} M \mathcal{T} )
= \mathrm{Tr} ( M^+ )$ valid for linear operators $M$. 
This means that the entire tensor $T^{(N)}$ vanishes
identically for $N$ odd, which is consistent with the eigenvalues
of the considered systems being even functions of the $\mathbf{k}$
vector, $E^{(n)}_s(-\mathbf{k}) = E^{(n)}_s(\mathbf{k})$.

Let us examine now the terms in the expansion (\ref{eq_tayfk}) that
are even in $\mathbf{k}$; we will employ the transformations given
by Eq.~(\ref{eq_trpall}).
For the reference term $\mathrm{Tr} (\sigma g)$, for an arbitrary 
element $(\alpha , \eta) \in \mathcal{P}_\mathrm{M}$, and with
abbreviation $\mathcal{D}(\alpha , \eta) = D$, we get:
\begin{equation}
\mathrm{Tr} (\sigma g) =
\mathrm{Tr} ( D \eta \sigma D^{-1} D g D^{-1} ) =
\eta \mathrm{Tr} ( \sigma g ) .
\label{eq_scn0}
\end{equation}
This means that for $\mathcal{P}_\mathrm{M}$ of category (a) or
(c), which contains elements $(\alpha , -1)$, the term
$\mathrm{Tr} (\sigma g)$ vanishes and there is no spin splitting
of the bands in the very center of the BZ.
For ferromagnets and ferrimagnets, featured by
$\mathcal{P}_\mathrm{M}$ of category (b), the eigenstates are
obviously spin split for all $\mathbf{k}$ points.
In the following, we thus confine ourselves to magnetic point
groups of categories (a) and (c), i.e., to groups with some
elements containing the operation of antisymmetry.

Let us further discuss in detail the shape of the tensor
$T^{(2)}_{\mu_1 \mu_2}$, Eq.~(\ref{eq_tayfk}).
For $(\alpha , \eta) \in \mathcal{P}_\mathrm{M}$ and with
abbreviation $\mathcal{D}(\alpha , \eta) = D$, we get:
\begin{equation}
T^{(2)}_{\mu_1 \mu_2} = \mathrm{Tr}
( D \eta \sigma D^{-1} D g D^{-1} W^{(2)}_{\mu_1 \mu_2} D g D^{-1} )
 = \eta \sum_{\nu_1 \nu_2} \alpha_{\mu_1 \nu_1} \alpha_{\mu_2 \nu_2}
\mathrm{Tr} ( \sigma g W^{(2)}_{\nu_1 \nu_2} g ) ,
\label{eq_scn2aux}
\end{equation}
so that the condition for each element
$(\alpha , \eta) \in \mathcal{P}_\mathrm{M}$ is
\begin{equation}
T^{(2)}_{\mu_1 \mu_2} = \eta
 \sum_{\nu_1 \nu_2} \alpha_{\mu_1 \nu_1} \alpha_{\mu_2 \nu_2}
 T^{(2)}_{\nu_1 \nu_2} .
\label{eq_scn2}
\end{equation}
However, this approach does not account explicitly
for the tensor symmetry,
$T^{(2)}_{\mu_1 \mu_2} = T^{(2)}_{\mu_2 \mu_1}$.
In order to include this property, one has to modify the condition
(\ref{eq_scn2}) to
\begin{equation}
T^{(2)}_{\mu_1 \mu_2} = \frac{1}{2} \eta
\sum_{\nu_1 \nu_2} ( \alpha_{\mu_1 \nu_1} \alpha_{\mu_2 \nu_2} +
\alpha_{\mu_2 \nu_1} \alpha_{\mu_1 \nu_2} ) T^{(2)}_{\nu_1 \nu_2} .
\label{eq_scn2sym}
\end{equation}
The validity of the last condition for all group elements leads
then to the final condition on the shape of the tensor
$T^{(2)}_{\mu_1 \mu_2}$ as
\begin{eqnarray}
T^{(2)}_{\mu_1 \mu_2} & = & \sum_{\nu_1 \nu_2}
Q^{(2)}_{\mu_1 \mu_2 , \nu_1 \nu_2} T^{(2)}_{\nu_1 \nu_2} ,
\nonumber\\
Q^{(2)}_{\mu_1 \mu_2 , \nu_1 \nu_2} & = &
\frac{1}{ 2 | \mathcal{P}_\mathrm{M} | }
\sum_{ (\alpha , \eta) }^{ \mathcal{P}_\mathrm{M} }
\eta ( \alpha_{\mu_1 \nu_1} \alpha_{\mu_2 \nu_2}
     + \alpha_{\mu_2 \nu_1} \alpha_{\mu_1 \nu_2} ) ,
\label{eq_scn2fin}
\end{eqnarray}
where the last sum runs over all elements $(\alpha , \eta)$
of the magnetic point group $\mathcal{P}_\mathrm{M}$ and where
$| \mathcal{P}_\mathrm{M} |$ denotes its order (number of group
elements).
It can be shown that the introduced superoperator
$Q^{(2)}_{\mu_1 \mu_2 , \nu_1 \nu_2}$ is a projector in a
9-dimensional vector space, i.e., it is symmetric,
$Q^{(2)}_{\mu_1 \mu_2 , \nu_1 \nu_2} =
 Q^{(2)}_{\nu_1 \nu_2 , \mu_1 \mu_2}$, and idempotent,
\begin{equation}
\sum_{\lambda_1 \lambda_2}
  Q^{(2)}_{\mu_1 \mu_2 , \lambda_1 \lambda_2}
  Q^{(2)}_{\lambda_1 \lambda_2 , \nu_1 \nu_2}
= Q^{(2)}_{\mu_1 \mu_2 , \nu_1 \nu_2} .
\label{eq_q2idemp}
\end{equation}
Consequently, the number $q^{(2)}$ of independent nonzero components
of the tensor $T^{(2)}_{\mu_1 \mu_2}$ can easily be obtained as the
trace of the projector $Q^{(2)}_{\mu_1 \mu_2 , \nu_1 \nu_2}$,
namely,
\begin{equation}
q^{(2)} = \sum_{\mu_1 \mu_2} Q^{(2)}_{\mu_1 \mu_2 , \mu_1 \mu_2} .
\label{eq_q2trace}
\end{equation}
The shape of the tensor $T^{(2)}_{\mu_1 \mu_2}$, given by 
Eq.~(\ref{eq_scn2fin}), was derived by considering only the
elements of the group $\mathcal{P}_\mathrm{M}$; it can easily be
shown that inclusion of both additional symmetries, mentioned
in the end of Section~\ref{ss_pes}, has no influence on the
obtained result.
This follows from the commutation of the spin operator $\sigma$
with operators $h$, $g$, and $W^{(N)}_{\mu_1 \mu_2}$, as well as
from the obvious modification of Eq.~(\ref{eq_trtnt}) for $N$ even.

The derived shape of the tensor $T^{(2)}_{\mu_1 \mu_2}$ is closely
related to spin conductivity.
The latter property, defined as the linear response of a spin current
to an external electric field, is usually quantified by a tensor
$\sigma^\lambda_{\mu_1 \mu_2}$, where the Cartesian index $\lambda$ 
refers to the spin polarization of the spin current, $\mu_1$ 
corresponds to the direction of the spin-current flow, and $\mu_2$ to
the direction of the electric field \cite{r_2015_skw, r_2021_hsv}.
In nonrelativistic collinear magnets, the two-current model of
electron transport is valid \cite{r_1967_cfp}, the original tensor
reduces to $\sigma^\lambda_{\mu_1 \mu_2} = n_\lambda
\tilde{\sigma}_{\mu_1 \mu_2}$, where $(n_x, n_y, n_z) = \mathbf{n}$
is a unit vector parallel to all magnetic moments, and the shape of
the tensor $\tilde{\sigma}_{\mu_1 \mu_2}$ coincides with that of
$T^{(2)}_{\mu_1 \mu_2}$, see Appendix \ref{app_sc}.
This fact points to a close relation between the spin splitting of
the electronic band structure and the spin conductivity, which is
one of the central properties in spintronics.

We turn finally to the case of a general even $N$.
In full analogy with Eq.~(\ref{eq_scn2}) for $N=2$, the condition
on the tensor $T^{(N)}_{\mu_1 \mu_2 \dots \mu_N}$ can be written
for each $(\alpha , \eta) \in \mathcal{P}_\mathrm{M}$ as
\begin{equation}
T^{(N)}_{\mu_1 \mu_2 \dots \mu_N} =
\eta \sum_{\nu_1 \nu_2 \dots \nu_N}
\alpha_{\mu_1 \nu_1} \alpha_{\mu_2 \nu_2} \dots
\alpha_{\mu_N \nu_N} T^{(N)}_{\nu_1 \nu_2 \dots \nu_N} .
\label{eq_scn}
\end{equation}
Explicit inclusion of the tensor symmetry (invariance of
$T^{(N)}_{\mu_1 \mu_2 \dots \mu_N}$ with respect to all permutations
of the indices) leads to a modified condition of the form
\begin{equation}
T^{(N)}_{\mu_1 \mu_2 \dots \mu_N} = \frac{1}{N!}
\eta \sum_{\nu_1 \nu_2 \dots \nu_N} \mathrm{per}
( \tilde{\alpha}^{\mu_1 \mu_2 \dots \mu_N ,
  \nu_1 \nu_2 \dots \nu_N} ) T^{(N)}_{\nu_1 \nu_2 \dots \nu_N} ,
\label{eq_scnsym}
\end{equation}
where the symbol $\mathrm{per} ( C )$ denotes the permanent
of a square matrix $C$ and where \\
$\tilde{\alpha}^{\mu_1 \mu_2 \dots \mu_N , \nu_1 \nu_2 \dots \nu_N}$
denotes a square $N \times N$ matrix with elements
\begin{equation}
\{ \tilde{\alpha}^{\mu_1 \mu_2 \dots \mu_N ,
\nu_1 \nu_2 \dots \nu_N} \}_{ij} = \alpha_{\mu_i \nu_j}
\quad \ \mathrm{for} \ \, i, j \in \{ 1 , 2, \dots , N \} .
\label{eq_tadef}
\end{equation}
The final condition on the tensor shape is:
\begin{eqnarray}
T^{(N)}_{\mu_1 \mu_2 \dots \mu_N} & = &
\sum_{\nu_1 \nu_2 \dots \nu_N}
Q^{(N)}_{\mu_1 \mu_2 \dots \mu_N , \nu_1 \nu_2 \dots \nu_N}
T^{(N)}_{\nu_1 \nu_2 \dots \nu_N} ,
\nonumber\\
Q^{(N)}_{\mu_1 \mu_2 \dots \mu_N , \nu_1 \nu_2 \dots \nu_N} & = &
\frac{1}{ N! \, | \mathcal{P}_\mathrm{M} | }
\sum_{ (\alpha , \eta) }^{ \mathcal{P}_\mathrm{M} } \eta
\mathrm{per} ( \tilde{\alpha}^{\mu_1 \mu_2 \dots \mu_N ,
  \nu_1 \nu_2 \dots \nu_N} ) ,
\label{eq_scnfin}
\end{eqnarray}
and the number $q^{(N)}$ of independent nonzero components of
the tensor $T^{(N)}_{\mu_1 \mu_2 \dots \mu_N}$ equals to
\begin{equation}
q^{(N)} = \sum_{\mu_1 \mu_2 \dots \mu_N}
Q^{(N)}_{\mu_1 \mu_2 \dots \mu_N , \mu_1 \mu_2 \dots \mu_N} .
\label{eq_qntrace}
\end{equation}
The last two equations represent the main result of this section.

Evaluation of the projection superoperator 
$Q^{(N)}_{\mu_1 \mu_2 \dots \mu_N , \nu_1 \nu_2 \dots \nu_N}$ for
selected groups $\mathcal{P}_\mathrm{M}$ was straightforward,
based on the known group elements $(\alpha , \eta)$ and rotation
matrices $\alpha = \{ \alpha_{\mu \nu} \}$.
The identification of nonvanishing components of the tensor
$T^{(N)}_{\mu_1 \mu_2 \dots \mu_N}$ and the linear dependences
among them were derived from the identification of nonzero rows
of the superoperator
$Q^{(N)}_{\mu_1 \mu_2 \dots \mu_N , \nu_1 \nu_2 \dots \nu_N}$
and from the linear dependences among them.
The success of this simple approach rests on the adopted orientation
of rotation axes and mirror planes of the considered point groups
with respect to the Cartesian coordinate system (for details, see
Appendix \ref{app_drsa}).
However, the number $q^{(N)}$ is insensitive to this orientation.

\section{Results and discussion\label{s_redi}}

\subsection{Magnetic point groups of selected
            systems\label{ss_mpgss}}

\begin{table}[htb]
\caption{Directions of magnetic moments ($\mathbf{n}$),
standard magnetic point groups (st-MPG), and modified magnetic point
groups (mod-MPG) with pseudoscalar spin for selected collinear
magnetic systems.
The parentheses at the group symbols contain the group orders.
\label{t_mpg}}
\begin{ruledtabular}
\begin{tabular}{lccc}
system & $\mathbf{n}$ & st-MPG & mod-MPG \\
\hline
Fe & (001)  & 4/mm$'$m$'$ (16) & m$\bar{3}$m (48) \\
Co & (0001) & 6/mm$'$m$'$ (24) & 6/mmm (24) \\
Ni & (111) & $\bar{3}$m$'$ (12) & m$\bar{3}$m (48) \\
FeO & (111) & $\bar{3}$m1$'$ (24) & $\bar{3}$m1$'$ (24) \\
Mn$_2$Au & (100), (110) & m$'$mm (8) & 4/m$'$mm (16) \\
RuO$_2$ & (001) & 4$'$/mm$'$m (16) & 4$'$/mm$'$m (16) \\  
RuO$_2$ & (100), (110) & m$'$m$'$m (8) & 4$'$/mm$'$m (16) \\  
MnTe & (11$\bar{2}$0) & mmm (8) & 6$'$/m$'$m$'$m (24) \\  
MnTe & (1$\bar{1}$00) & m$'$m$'$m (8) & 6$'$/m$'$m$'$m (24) \\  
\end{tabular}
\end{ruledtabular}
\end{table}

As mentioned in Section \ref{ss_pes}, the replacement
of the original vector spin, exchange field, and
local magnetic moments by their pseudoscalar counterparts
(accompanied also by switching off spin-orbit interaction) leads to
modified magnetic point groups for real systems \cite{r_sumampg}.
As an illustration, we present in Table~\ref{t_mpg} the standard
and modified magnetic point groups $\mathcal{P}_\mathrm{M}$ for
three elemental ferromagnets (bcc Fe, hcp Co, and fcc Ni)
and four binary antiferromagnetic compounds: 
FeO with a rocksalt structure \cite{r_1958_wlr},
Mn$_2$Au with a body-centered tetragonal (bct) structure 
\cite{r_2013_bcm, r_2018_bst},
RuO$_2$ with a rutile structure \cite{r_2019_zsr, r_2020_sgj},
and MnTe with a hexagonal structure \cite{r_2017_krg}.
The selected antiferromagnets are featured by simple magnetic
structures, with one formula unit per magnetic unit cell for Mn$_2$Au
\cite{r_2013_bcm, r_2018_bst}, while two formula units form one
magnetic unit cell in FeO \cite{r_1958_wlr},
RuO$_2$ \cite{r_2019_zsr}, and MnTe \cite{r_2017_krg}; for all
compounds, the positions of nonmagnetic atoms are taken into account.
In the standard treatment, the resulting symmetry depends on the
direction $\mathbf{n}$ of magnetic moments, whereas the results of
the modified approach are insensitive to this direction.
For ferromagnets listed in Table~\ref{t_mpg}, 
the standard magnetic point groups belong to
category (c), while the modified ones belong to category (b), being
identical with the crystallographic point groups of the underlying
cubic (Fe, Ni) and hexagonal (Co) lattices.
Different situations are found for antiferromagnets.
For FeO, both groups are identical, belonging to category (a).
For RuO$_2$ with magnetic moments along (001) direction (the fourfold
axis), both groups belong to category (c) and they represent two
different versions of the group 4$'$/mm$'$m [the primed reflections
are on the (110) and (100) planes in the standard and modified
$\mathcal{P}_\mathrm{M}$, respectively].
For all other systems, the modified groups belong to category (c) as
well, but they differ explicitly from the standard ones.
Moreover, no direct group-subgroup relation could be found in these
cases between the modified and standard $\mathcal{P}_\mathrm{M}$.
Nevertheless, one can observe in Table~\ref{t_mpg} that the group
order of the standard $\mathcal{P}_\mathrm{M}$ divides that of the
modified $\mathcal{P}_\mathrm{M}$ in all cases studied,
see Ref.~\onlinecite{r_sumampg} for more details.

The modification of the magnetic point groups leads naturally to a
modification of the magnetic space groups.
As an example, we mention the antiferromagnetic MnF$_2$ compound
with a rutile structure and with Mn moments pointing along the
tetragonal axis \cite{r_2020_ywl}; this system is equivalent
to RuO$_2$ with Ru moments along (001) direction.
Its standard magnetic space group (for the system with spin-orbit
interaction) is P4$'_2$/mnm$'$ and the modified group is
P4$'_2$/mn$'$m \cite{r_2020_ywl}, in agreement with the two versions
of the point group 4$'$/mm$'$m of RuO$_2$.
A more detailed discussion of the space groups goes beyond
the scope of the present study.

\subsection{Classification of collinear nonrelativistic
            magnets\label{ss_ccnm}}

Inspection of the derived general formula for $N$ even,
Eq.~(\ref{eq_scnfin}), reveals that the superoperator $Q^{(N)}$
depends only on the Laue class of $\mathcal{P}_\mathrm{M}$ (the
magnetic Laue group is obtained by adding space inversion to all
elements of the magnetic point group $\mathcal{P}_\mathrm{M}$
\cite{r_2015_skw}).
This resembles the case of certain tensors, such as the conductivity
tensor and the tensor of thermoelectric coefficients
\cite{r_1966_whk, r_2015_skw},
and it simplifies the analysis of possible shapes of the tensors
$T^{(N)}$ substantially.
However, it should be noted that some of the magnetic point groups
of category (c) belong to the Laue class of category (a); this
happens if (and only if) the $\mathcal{P}_\mathrm{M}$ contains the
combination of space inversion and of the operation of antisymmetry.
Further inspection of Eq.~(\ref{eq_scnfin}) proves that for a
particular $\mathcal{P}_\mathrm{M}$ (or its Laue
class) of category (a), the superoperators $Q^{(N)}$ and the
resulting tensors $T^{(N)}$ vanish identically for all $N$. 
The evaluation of Eqs.~(\ref{eq_scnfin}) and (\ref{eq_qntrace})
has thus to be performed only for magnetic Laue groups of
category (c); the total number of these nontrivial magnetic Laue
groups amounts to ten.

\begin{table}[htb]
\caption{Nontrivial magnetic Laue groups (MLG) (in parenthesis
the subgroup $\mathcal{S}$ of all elements without the operation
of antisymmetry),
the lowest rank $N$ of a nonvanishing symmetric tensor $T^{(N)}$,
the nature B or P of the leading term in the Taylor expansion of
$F(\mathbf{k})$, and the number $q^{(N)}$ of
independent nonzero components of the tensor $T^{(N)}$.
\label{t_cap}}
\begin{ruledtabular}
\begin{tabular}{lccc}
MLG ($\mathcal{S}$) & $N$ & B/P & $q^{(N)}$ \\
\hline
m$'$m$'$m (2/m) & 2 & P & 1 \\
2$'$/m$'$ ($\bar{1}$) & 2 & B & 2 \\
4$'$/m (2/m) & 2 & P & 2 \\
4$'$/mm$'$m (mmm) & 2 & P & 1 \\
$\bar{3}$m$'$ ($\bar{3}$) & 4 & B & 1 \\
4/mm$'$m$'$ (4/m) & 4 & P & 1 \\
6$'$/m$'$ ($\bar{3}$) & 4 & B & 2 \\
6$'$/m$'$m$'$m ($\bar{3}$m) & 4 & B & 1 \\
6/mm$'$m$'$ (6/m) & 6 & P & 1 \\
m$\bar{3}$m$'$ (m$\bar{3}$) & 6 & B & 1 \\
\end{tabular}
\end{ruledtabular}
\end{table}

Our results are summarized in Table~\ref{t_cap}.
For each magnetic Laue group, the lowest rank $N$ of a 
nonvanishing tensor $T^{(N)}_{\mu_1 \mu_2 \dots \mu_N}$ is given
together with the number $q^{(N)}$ of its independent nonzero
components.
All these nonzero tensor components are listed explicitly in
Appendix \ref{app_drsa}; the symbols B (bulk) and P (planar)
in Table~\ref{t_cap} indicate which of the components $k_x$, $k_y$,
and $k_z$ of the $\mathbf{k}$ vector enter the leading term in the
Taylor expansion (\ref{eq_tayfk}) of $F(\mathbf{k})$.
The symbol P refers to the cases where only two components in
directions perpendicular to a prominent direction of the group are
present, while the symbol B denotes all other cases.
The most important observation is the fact that for each magnetic
Laue group of category (c), a nonvanishing tensor $T^{(N)}$
exists, which in turn proves the presence of spin splitting in
a neighborhood of the BZ center.

Let us discuss briefly the four antiferromagnets (FeO, Mn$_2$Au,
RuO$_2$, MnTe) mentioned in Section \ref{ss_mpgss}, see also
Table~\ref{t_mpg}.
The modified $\mathcal{P}_\mathrm{M}$ of FeO ($\bar{3}$m1$'$)
belongs to category (a), incompatible with spin splitting.
The modified $\mathcal{P}_\mathrm{M}$ of RuO$_2$ (4$'$/mm$'$m) and
that of MnTe (6$'$/m$'$m$'$m) are Laue groups of category (c),
compatible with spin splitting.
The modified $\mathcal{P}_\mathrm{M}$ of Mn$_2$Au (4/m$'$mm) is of
category (c); however, its Laue class (4/mmm1$'$) is of
category (a), which does not support spin splitting.

The present identification of a broad pool of ten nontrivial
magnetic Laue groups, yielding the spin splitting of energy bands
in antiferromagnets, is in full agreement with ample occurrence of
this phenomenon \cite{r_2016_non, r_2019_ahl, r_2019_hyk, r_2019_nhk,
r_2021_ywl_m, r_2021_mkj, r_2021_ssj}.
Moreover, a closer look at the results in Table~\ref{t_cap} reveals
a remarkable similarity with a classification scheme of
altermagnets obtained by \v{S}mejkal et al. \cite{r_2021_ssj}.
The approach developed by the authors of Ref.~\onlinecite{r_2021_ssj}
is based on spin point groups, on eigenvalues of model
$\mathbf{k} \cdot \mathbf{p}$ Hamiltonians, and on an
orbital-harmonic representation \cite{r_2020_hyk}.
The different ten altermagnetic cases, summarized in Table I of
Ref.~\onlinecite{r_2021_ssj}, are featured by the spin Laue group,
a spin winding number $W$ ($W = 2, 4, 6$), and the B/P symbol. 
A unique one to one mapping between the ten cases of \v{S}mejkal et
al.\ and those in Table~\ref{t_cap} can be found after identification
of $N$ with the spin winding number $W$ and by comparing the B/P
symbols, the parent crystallographic point groups, and the 
subgroups of index two attached to the magnetic/spin Laue groups. 
This mapping is further corroborated by the resulting
$\mathbf{k}$-dependent functions: $F(\mathbf{k})$,
Eq.~(\ref{eq_tayfk}), and its eigenvalue-based counterpart
\cite{r_2021_ssj}, see Appendix \ref{app_drsa}.

The similarity of the results of both classification schemes
deserves a brief comment.
In the approach using the spin point groups, the reversal of local
magnetic moments in antiferromagnets is achieved by the $\pi$
rotation in the spin space around an axis perpendicular to the
moment direction \cite{r_2021_ssj}.
In the pseudoscalar-spin approach, the local moment reversal is
owing to the operation of antisymmetry (spin-channel interchange)
present in the modified magnetic groups.
The latter approach leads then to a very simple classification
of nonrelativistic collinear magnets:
the ferro- and ferrimagnets (including the compensated ones) with
different spin-up and spin-down band structures are characterized
by the magnetic point group of category (b), the usual
antiferromagnets without spin-split electronic structure possess
the magnetic Laue group of category (a), and the antiferromagnets
with spin splitting (altermagnets) are featured by the magnetic
Laue group of category (c).
These three categories of magnetic point and Laue groups proved 
very useful for understanding the transport phenomena in magnetic
materials since the 1960's; one might expect that they will also be
helpful in the field of antiferromagnets with momentum-dependent
spin splitting.

As an example, let us consider the spin conductivity introduced
in Section \ref{ss_sast}.
The shape of the spin-conductivity tensor
$\tilde{\sigma}_{\mu_1 \mu_2}$ coincides with that of the tensor
$T^{(2)}_{\mu_1 \mu_2}$.
According to Table~\ref{t_cap}, this tensor is nonzero only for
four magnetic Laue classes, namely for m$'$m$'$m, 2$'$/m$'$,
4$'$/m, and 4$'$/mm$'$m.
This result explains different sources of the calculated spin
conductivities of hexagonal MnTe and tetragonal RuO$_2$ systems
\cite{r_2021_hsv}: in MnTe (modified $\mathcal{P}_\mathrm{M}$
6$'$/m$'$m$'$m) it is caused solely by spin-orbit interaction,
whereas in RuO$_2$ (modified $\mathcal{P}_\mathrm{M}$ 4$'$/mm$'$m)
it is induced primarily by the anisotropic spin-split bands.
The anisotropy of RuO$_2$ can easily be understood by inspecting
the subgroup $\mathcal{S}$ of all elements without the operation of
antisymmetry, which is the orthorhombic group mmm with mirror planes
(001), (110), and (1$\bar{1}$0).
Consequently, the conductivities in each spin channel are different
along the (110) and (1$\bar{1}$0) directions, which (together with
the spin-channel interchange accompanying the rotation by $\pi/2$
around $z$ axis present in $\mathcal{P}_\mathrm{M}$) leads to the
resulting nonzero spin conductivity \cite{r_2019_ahl, r_2021_hsv}.

The obtained classification scheme is also compatible with the
recently formulated criteria for spin splitting in antiferromagnets,
based on magnetic space groups \cite{r_2020_ywl, r_2021_ywl_m}.
All magnetic space groups $\mathcal{G}_\mathrm{M}$ can be divided
into four types \cite{r_2010_bc} which correspond to the three
categories of the magnetic point groups $\mathcal{P}_\mathrm{M}$
derived from the $\mathcal{G}_\mathrm{M}$ as follows.
A group $\mathcal{G}_\mathrm{M}$ of type I does not involve the
operation of antisymmetry at all (neither as a separate element nor
in a combination with a spatial operation);
its $\mathcal{P}_\mathrm{M}$ belongs to category (b).
A group $\mathcal{G}_\mathrm{M}$ of type II contains the pure
operation of antisymmetry as a group element;
its $\mathcal{P}_\mathrm{M}$ belongs to category (a).
A group $\mathcal{G}_\mathrm{M}$ of type III contains the
operation of antisymmetry only in a combination with a nontrivial
rotation (combined optionally with a translation);
its $\mathcal{P}_\mathrm{M}$ belongs to category (c).
For this type, further partitioning can be done which is equivalent
to the two categories [(a) or (c)] relevant for the Laue class of
the $\mathcal{P}_\mathrm{M}$ of category (c).
A group $\mathcal{G}_\mathrm{M}$ of type IV contains the operation
of antisymmetry in a combination with a nonprimitive translation;
its $\mathcal{P}_\mathrm{M}$ belongs to category (a).
However, in application of both approaches to a particular system,
different nature of electron spin (vector or pseudoscalar) should
also be taken into account.
As an illustrating example, we consider the antiferromagnetic NiO
system with a perturbed rocksalt structure in which each oxygen
(111) plane is displaced slightly along (111) direction towards the
nearest nickel (111) plane with positive magnetic moments
\cite{r_2021_ywl_b}.
For Ni moments oriented along (11$\bar{2}$) direction, the standard
$\mathcal{G}_\mathrm{M}$ is C2$'$/m$'$ which is of type III and which
leads to the spin splitting of eigenvalues essentially throughout
the whole BZ.
Within the pseudoscalar-spin approach, the modified 
$\mathcal{P}_\mathrm{M}$ is $\bar{3}$m which belongs to category (b)
leading thus to the same kind of spin splitting.
This can easily be understood in terms of ferrimagnetism: nickel
atoms with opposite signs of local moments behave (from a viewpoint
of symmetry) as two chemically different species due to the adopted
displacements of oxygen atoms. 
The system can thus be treated as a nearly compensated ferrimagnet
with different band structures in spin-up and spin-down channels
which explains the resulting spin splitting.
A more detailed comparison of the approach based on
$\mathcal{G}_\mathrm{M}$ \cite{r_2020_ywl, r_2021_ywl_m} and the
current one employing $\mathcal{P}_\mathrm{M}$ goes beyond the scope
of this work.

\subsection{Spin splitting in a model
            antiferromagnet\label{ss_ssma}}

In this section, we discuss briefly the physical mechanisms
behind the spin splitting of eigenvalues in collinear
antiferromagnets. 
Since the only known monatomic collinear antiferromagnet is
chromium on a bcc lattice, which however forms a spin-density wave
with a wavelength incommensurate with the bcc lattice parameter
\cite{r_1988_ef}, one can conclude that nonmagnetic atoms play an
important role for collinear antiferromagnets with perfect
translation invariance.
The effect of the nonmagnetic atoms is manifold.
First, they are responsible for stabilization of the geometric
structure of the systems.
Second, they often lead to the formation of the local magnetic
moments and to their antiferromagnetic exchange coupling.
Finally, the nonmagnetic atoms create local electric crystal fields
around the magnetic atoms, which in combination with spin-group
symmetries of the one-electron Hamiltonian give rise to the spin
splitting \cite{r_2021_ssj}.
The important role of nonmagnetic atoms has also been proved in a
recent study of NiO with a rocksalt structure, in which small
displacements of oxygen atoms were introduced, which generated a
pronounced spin splitting \cite{r_2021_ywl_b}.

Let us assess relative importance of the nonmagnetic atoms and of
the group symmetry using an example of the antiferromagnetic
KRu$_4$O$_8$ compound with a bct structure, which exhibits the
spin splitting \cite{r_2021_ssj}.
The potassium atoms occupy the Wyckoff 2(b) positions of the space
group (space group I4/m, No.~87), while the ruthenium atoms
as well as both kinds of oxygen atoms occupy the Wyckoff 8(h)
positions \cite{r_1978_wh}.
The three fundamental vectors of the Bravais bct lattice are
$\mathbf{a}_1 = (a, 0, 0)$, $\mathbf{a}_2 = (0, a, 0)$, and
$\mathbf{a}_3 = (a/2, a/2, c/2)$, where $a$ and $c$ are the
bct lattice parameters.
The basis vectors of Ru atoms are $\mathbf{B}_1 = (ua, va, 0)$,
$\mathbf{B}_2 = (-va, ua, 0)$,  $\mathbf{B}_3 = - \mathbf{B}_1$, and
$\mathbf{B}_4 = - \mathbf{B}_2$, where $u$ and $v$ are dimensionless
atomic coordinates.
The local magnetic moments of Ru atoms at $\mathbf{B}_1$ and
$\mathbf{B}_3$ are identical, being opposite to those of Ru atoms
at $\mathbf{B}_2$ and $\mathbf{B}_4$.
The magnetic point group (with pseudoscalar spin) of the whole
system is 4$'$/m, which is compatible with existence of the spin
splitting according to Table~\ref{t_cap}.
However, the same magnetic point group is also obtained for
a hypothetical four-site bct system derived from KRu$_4$O$_8$ by
removing all nonmagnetic atoms (K, O), keeping thus only the
magnetic Ru atoms with their antiferromagnetic structure.
This indicates that the splin splitting might be obtained even
without any nonmagnetic atoms in this case.

\begin{figure}[htb]
\begin{center}
\includegraphics[width=0.50\textwidth]{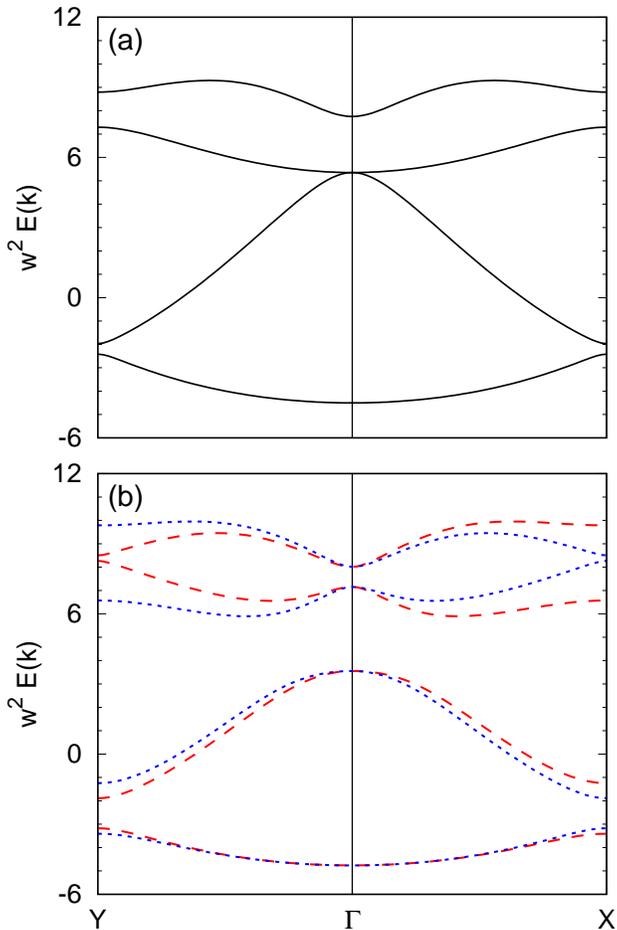}
\end{center}
\caption{
The band structure of the four-site model on a bct lattice:
(a) in the nonmagnetic state, (b) in the antiferromagnetic state.
In panel (b), the dotted and dashed curves correspond to the
two spin channels, $s=1$ and $s=-1$, respectively.
\label{f_bct}}
\end{figure}

In order to verify this idea on a very simple model, we performed
bandstructure calculations for this four-site bct model using
the linear muffin-tin orbital (LMTO) method in the atomic sphere
approximation \cite{r_1986_apj}, in which the angular-momentum
cutoff was set to $\ell_\mathrm{max} = 0$, corresponding thus
to a single orbital per atom.
The LMTO potential parameter $\Delta$, which controls the bandwidth,
was set to $\Delta = 1.8 w^{-2}$, where $w$ denotes the Wigner-Seitz
radius of the lattice, and the dimensionless LMTO potential
parameter was taken $\gamma = 0.4$, which is a typical value for
$s$ orbitals of transition metals.
The LMTO potential parameter $C$, which controls the position of
the bands on energy scale, was set zero for a nonmagnetic system,
whereas exchange-split values of $C = \pm \Delta$ were used to
simulate the antiferromagnetic order.
The geometric structure of the model is defined by $u=0.27$,
$v=0.08$, and $c/a = 0.4$ (these values were chosen in order to
achieve good space filling by the atomic spheres).

The resulting band structures are displayed in Fig.~\ref{f_bct}
along the $\mathrm{Y} - \Gamma - \mathrm{X}$ path in the bct BZ,
i.e., for $k_z = 0$, $k_y = | k_x |$, and $- \pi \le a k_x \le \pi$.
One can observe clearly the spin splitting in the antiferromagnetic
state, in qualitative agreement with that found in the realistic
band structure of the KRu$_4$O$_8$ compound \cite{r_2021_ssj}.
A similar result has recently been obtained for a single-orbital
four-site antiferromagnetic model without nonmagnetic atoms, derived
from a pyrochlore structure \cite{r_2019_hyk} and characterized by
the magnetic point group 4$'$/mm$'$m.
These results prove that the nontrivial magnetic Laue group of the
system is the most important prerequisite for the appearance of spin
splitting; other features present in real materials, such as, e.g.,
the local electric fields and magnetic exchange interactions due
to the nonmagnetic atoms, or the true orbital structure of the
magnetic transition-metal atoms, are less essential in this respect
(despite their primary importance for the realistic band structure
and the splitting strength).

\section{Conclusions\label{s_conc}}

We have introduced the concept of a pseudoscalar electron spin
appropriate for a theoretical treatment of the electronic
structure of nonrelativistic collinear magnets.
The substitution of the original vector spin by the pseudoscalar
spin brings about a modification of magnetic groups of crystalline
systems.
We have defined an infinite-dimensional representation of the
modified magnetic point groups, which enabled us to avoid any
approximations in solving the Hamiltonian eigenvalue problem.
This representation is single-valued and unitary,
which might be beneficial in future extensions of the formalism.

The developed theory was used for an analysis of spin splitting
of electron states in antiferromagnets near the center of the
Brillouin zone.
Our results provide an alternative view on the recently introduced
altermagnetic systems \cite{r_2021_ssj}; their different classes were
identified unambiguously with the nontrivial magnetic Laue classes
that are relevant for shape restrictions of various transport tensor
quantities.
As a consequence, the spin conductivity induced by spin-split bands
in certain antiferromagnets has been ascribed to four specific
magnetic Laue classes.
A brief discussion of a model antiferromagnet without nonmagnetic
atoms revealed that the point-group symmetry of the system represents
the key factor for existence of the spin splitting while the
nonmagnetic atoms influence only the splitting magnitude.

The present work was confined to magnetic point groups; it can be,
together with recent studies based on magnetic space groups
\cite{r_2020_ywl, r_2021_ywl_m, r_2021_ywl_b} and on spin point
groups \cite{r_2021_ssj}, considered as one of the starting points
on a way towards a complete symmetry analysis of 
transport properties and electron states
in collinear nonrelativistic magnets, which should inevitably
include spin space groups \cite{r_2022_llh}.
This topic has to be left for the future. 



\appendix

\section{Hamiltonian and transformation of
         operators\label{app_hto}}

The Hamiltonian (\ref{eq_hrdef}) is represented in the basis
of orthonormal vectors $| \mathbf{G} s \rangle = 
| \mathbf{G} \rangle \otimes | s \rangle$, where the spatial
part for a given $\mathbf{k}$ point is defined by
\begin{equation}
\langle \mathbf{r} | \mathbf{G} \rangle = \Omega^{-1/2}
\exp[ i ( \mathbf{k} - \mathbf{G} ) \cdot \mathbf{r} ] ,
\label{eq_app_rg}
\end{equation}
where $\Omega$ denotes the volume of the primitive cell in the real
space.
The full Hamiltonian matrix in this basis is
\begin{equation}
\langle \mathbf{G} s | \tilde{H}(\mathbf{k})
 | \mathbf{G}' s' \rangle =
[ ( \mathbf{k} - \mathbf{G} )^2 \delta_{\mathbf{G} \mathbf{G}'}
 + \tilde{V}_s(\mathbf{G} - \mathbf{G}') ] \delta_{ss'} ,
\label{eq_app_hkmat}
\end{equation}
where we introduced Fourier coefficients
$\tilde{V}_s(\mathbf{G})$ of the potentials $V_s(\mathbf{r})$,
so that
\begin{equation}
V_s(\mathbf{r}) = \sum_\mathbf{G} \tilde{V}_s(\mathbf{G})
\exp ( - i \mathbf{G} \cdot \mathbf{r} ) .
\label{eq_app_fcvs}
\end{equation}
Consequently, matrix elements of the operators $h$, $J_\mu$, and
$L_{\mu_1 \mu_2}$, which define the full $\mathbf{k}$-dependence
of $\tilde{H}(\mathbf{k})$, Eq.~(\ref{eq_hktayl}), are given by
\begin{eqnarray}
\langle \mathbf{G} s | h | \mathbf{G}' s' \rangle & = &
[ \mathbf{G}^2 \delta_{\mathbf{G} \mathbf{G}'}
+ \tilde{V}_s(\mathbf{G} - \mathbf{G}') ] \delta_{ss'} ,
\nonumber\\
\langle \mathbf{G} s | J_\mu | \mathbf{G}' s' \rangle & = &
- 2 G_\mu \delta_{\mathbf{G} \mathbf{G}'} \delta_{ss'} ,
\nonumber\\
\langle \mathbf{G} s | L_{\mu_1 \mu_2} | \mathbf{G}' s' \rangle
 & = &
\delta_{\mu_1 \mu_2} \delta_{\mathbf{G} \mathbf{G}'} \delta_{ss'} .
\label{eq_app_hjl}
\end{eqnarray}
The last relation implies that $L_{\mu_1 \mu_2} =
I \delta_{\mu_1 \mu_2}$, where $I$ is the unit operator.
We also mention the matrix elements of the spin operator $\sigma$,
which reduce to
\begin{equation}
\langle \mathbf{G} s | \sigma | \mathbf{G}' s' \rangle =
s \delta_{\mathbf{G} \mathbf{G}'} \delta_{ss'} .
\label{eq_app_sigma}
\end{equation}
Note that Eq.~(\ref{eq_app_hkmat}) represents a starting point for
accurate eigenvalues of the Hamiltonian (\ref{eq_hrdef}), provided
that the basis set $\{ | \mathbf{G} s \rangle \}$ is not truncated.

Let us prove now the basic property (\ref{eq_repbr}) of the 
representation $\mathcal{D}(\alpha, \eta)$ of the magnetic point
group $\mathcal{P}_\mathrm{M}$, defined by Eq.~(\ref{eq_drepdef}).
If we denote $(\alpha_1, \eta_1) (\alpha_2, \eta_2) =
(\alpha_1 \alpha_2, \eta_1 \eta_2) \equiv (\alpha_3, \eta_3)$,
then the corresponding translations
$\mathbf{t}_j, j \in \{ 1,2,3 \}$, entering Eq.~(\ref{eq_mpgdef}),
satisfy $\mathbf{t}_3 = \alpha_1 \mathbf{t}_2 + \mathbf{t}_1$. 
We get for an arbitrary basis vector $| \mathbf{G} s \rangle$:
\begin{eqnarray}
&& \mathcal{D}(\alpha_1, \eta_1)
\mathcal{D}(\alpha_2, \eta_2) | \mathbf{G} s \rangle
 = \mathcal{D}(\alpha_1, \eta_1)
| \alpha_2 \mathbf{G}, \eta_2 s \rangle
\exp( i \alpha_2 \mathbf{G} \cdot \mathbf{t}_2 ) 
\nonumber\\
&& \quad = | \alpha_1 \alpha_2 \mathbf{G} ,
 \eta_1 \eta_2 s \rangle
\exp ( i \alpha_1 \alpha_2 \mathbf{G}
 \cdot \mathbf{t}_1 )
\exp ( i \alpha_2 \mathbf{G} \cdot \mathbf{t}_2 )
\nonumber\\
&& \quad = | \alpha_1 \alpha_2 \mathbf{G} ,
 \eta_1 \eta_2 s \rangle
\exp [ i \alpha_1 \alpha_2 \mathbf{G} \cdot
( \mathbf{t}_1 + \alpha_1 \mathbf{t}_2 ) ]
\nonumber\\
&& \quad = | \alpha_3 \mathbf{G}, \eta_3 s \rangle
\exp( i \alpha_3 \mathbf{G} \cdot \mathbf{t}_3 )
 = \mathcal{D}(\alpha_3 , \eta_3 ) | \mathbf{G} s \rangle .
\label{eq_app_d1d2}
\end{eqnarray}
This completes the proof of Eq.~(\ref{eq_repbr}).

Let us turn now to the transformation of relevant operators, as
summarized in Eq.~(\ref{eq_trpall}).
We start with the reference Hamiltonian $h$.
We evaluate $\mathcal{D}(\alpha, \eta) h$ and
$h \mathcal{D}(\alpha, \eta)$ for $(\alpha, \eta) \in
\mathcal{P}_\mathrm{M}$ and compare the results.
We get for an arbitrary basis vector $| \mathbf{G}' s \rangle$:
\begin{eqnarray}
&& \mathcal{D}(\alpha, \eta) h | \mathbf{G}' s \rangle = 
\mathcal{D}(\alpha, \eta) \sum_{\mathbf{G}'' }
 | \mathbf{G}'' s \rangle
\langle \mathbf{G}'' s | h | \mathbf{G}' s \rangle
\nonumber\\
&& \quad = \sum_{\mathbf{G}'' }
 | \alpha \mathbf{G}'', \eta s \rangle
\exp( i \alpha \mathbf{G}'' \cdot \mathbf{t} )
[ (\mathbf{G}')^2 \delta_{\mathbf{G}'' \mathbf{G}'}
+ \tilde{V}_s(\mathbf{G}'' - \mathbf{G}') ]
\nonumber\\
&& \quad = \sum_\mathbf{G} | \mathbf{G}, \eta s \rangle
\exp( i \mathbf{G} \cdot \mathbf{t} )
[ (\mathbf{G}')^2 \delta_{\mathbf{G}, \alpha \mathbf{G}' }
+ \tilde{V}_s(\alpha^{-1} \mathbf{G} - \mathbf{G}') ] ,
\label{eq_app_dh}
\end{eqnarray}
where we replaced the summation lattice vector
$\mathbf{G}''$ by $\mathbf{G} = \alpha \mathbf{G}''$.
Similarly, we get:
\begin{eqnarray}
&& h \mathcal{D}(\alpha, \eta) | \mathbf{G}' s \rangle
 = h | \alpha \mathbf{G}' , \eta s \rangle
\exp ( i \alpha \mathbf{G}' \cdot \mathbf{t} )
\nonumber\\
&& \quad = \sum_\mathbf{G} | \mathbf{G}, \eta s \rangle
\langle \mathbf{G}, \eta s | h | \alpha \mathbf{G}' ,
\eta s \rangle
\exp ( i \alpha \mathbf{G}' \cdot \mathbf{t} )
\nonumber\\
&& \quad = \sum_\mathbf{G} | \mathbf{G}, \eta s \rangle
\exp ( i \alpha \mathbf{G}' \cdot \mathbf{t} )
[ \mathbf{G}^2 \delta_{\mathbf{G}, \alpha \mathbf{G}' }
+ \tilde{V}_{\eta s}(\mathbf{G} - \alpha \mathbf{G}') ] .
\label{eq_app_hd}
\end{eqnarray}
The difference of both results yields:
\begin{eqnarray}
[ \mathcal{D}(\alpha, \eta) h - h \mathcal{D}(\alpha, \eta) ]
 | \mathbf{G}' s \rangle & = &
\sum_\mathbf{G} | \mathbf{G}, \eta s \rangle
\left[ \exp( i \mathbf{G} \cdot \mathbf{t} )
\tilde{V}_s(\alpha^{-1} \mathbf{G} - \mathbf{G}')
\right.
\nonumber\\
 & & \left. \qquad \quad \
{} - \exp ( i \alpha \mathbf{G}' \cdot \mathbf{t} )
\tilde{V}_{\eta s}(\mathbf{G} - \alpha \mathbf{G}') \right] .
\label{eq_app_dhmhd}
\end{eqnarray}
After application of an auxiliary identity (valid for all
reciprocal lattice vectors $\mathbf{G}$)
\begin{equation}
\tilde{V}_s(\mathbf{G}) =
\exp ( - i \alpha \mathbf{G} \cdot \mathbf{t} )
\tilde{V}_{\eta s}(\alpha \mathbf{G})
\label{eq_app_auxid}
\end{equation}
with $\mathbf{G}$ replaced by
$(\alpha^{-1} \mathbf{G} - \mathbf{G}')$, we get finally
\begin{equation}
\mathcal{D}(\alpha, \eta) h - h \mathcal{D}(\alpha, \eta) = 0 ,
\label{eq_app_hinv}
\end{equation}
which proves the invariance of the Hamiltonian $h$,
Eq.~(\ref{eq_trpall}).
The auxiliary identity (\ref{eq_app_auxid}) follows from
Eq.~(\ref{eq_mpgdef}) combined with the Fourier expansion
(\ref{eq_app_fcvs}):
\begin{eqnarray}
\sum_\mathbf{G} \tilde{V}_s(\mathbf{G})
\exp ( - i \mathbf{G} \cdot \mathbf{r} ) & = &
\sum_{\mathbf{G}'} \tilde{V}_{\eta s}(\mathbf{G}')
\exp [ - i \mathbf{G}' \cdot ( \alpha \mathbf{r} + \mathbf{t} ) ]
\nonumber\\
 & = & \sum_\mathbf{G} \tilde{V}_{\eta s}(\alpha \mathbf{G})
\exp [ - i \alpha \mathbf{G} \cdot ( \alpha \mathbf{r}
 + \mathbf{t} ) ] 
\nonumber\\
 & = & \sum_\mathbf{G} \tilde{V}_{\eta s}(\alpha \mathbf{G})
\exp ( - i \alpha \mathbf{G} \cdot \mathbf{t} )
\exp ( - i \mathbf{G} \cdot \mathbf{r} ) .
\label{eq_app_prai}
\end{eqnarray}
The comparison of coefficients at
$\exp ( - i \mathbf{G} \cdot \mathbf{r} )$ on both sides of this
relation yields the identity (\ref{eq_app_auxid}).

The transformation of the other operators can be obtained in a
similar way.
As an example, let us consider the velocities $J_\mu$.
We get for an arbitrary basis vector $| \mathbf{G} s \rangle$:
\begin{equation}
\mathcal{D}(\alpha, \eta) J_\mu | \mathbf{G} s \rangle = 
\mathcal{D}(\alpha, \eta) (-2 G_\mu) | \mathbf{G} s \rangle
 = -2 G_\mu | \alpha \mathbf{G} , \eta s \rangle
\exp ( i \alpha \mathbf{G} \cdot \mathbf{t} ) ,
\label{eq_app_djmu}
\end{equation}
and
\begin{eqnarray}
J_\mu \mathcal{D}(\alpha, \eta) | \mathbf{G} s \rangle & = &
J_\mu | \alpha \mathbf{G} , \eta s \rangle
\exp ( i \alpha \mathbf{G} \cdot \mathbf{t} )
 = -2 (\alpha \mathbf{G})_\mu
| \alpha \mathbf{G} , \eta s \rangle
\exp ( i \alpha \mathbf{G} \cdot \mathbf{t} )
\nonumber\\
 & = & -2 \sum_\nu \alpha_{\mu \nu} G_\nu
| \alpha \mathbf{G} , \eta s \rangle
\exp ( i \alpha \mathbf{G} \cdot \mathbf{t} ) .
\label{eq_app_jmud}
\end{eqnarray}
This means that
\begin{equation}
J_\mu \mathcal{D}(\alpha, \eta) =
\sum_\nu \alpha_{\mu \nu} \mathcal{D}(\alpha, \eta) J_\nu =
\mathcal{D}(\alpha, \eta) \sum_\nu \alpha_{\mu \nu} J_\nu ,
\label{eq_app_jtran}
\end{equation}
from which the transformation of $J_\mu$, Eq.~(\ref{eq_trpall}),
follows immediately.

\section{Spin conductivity\label{app_sc}}

The spin-conductivity tensor $\sigma^\lambda_{\mu_1 \mu_2}$ of a
nonrelativistic collinear magnet can be written according to a
general formula for the static linear response of noninteracting
electron systems \cite{r_1971_blb, r_2001_cb} at zero temperature as
\begin{eqnarray}
\sigma^\lambda_{\mu_1 \mu_2} & = &
-2c \int^{E_\mathrm{F}}_{-\infty} dE \overline{\mathrm{Tr}}
\left\{ \sigma^\lambda p_{\mu_1} Z'(E_+) p_{\mu_2}
[ Z(E_+) - Z(E_-) ] \right.
\nonumber\\
 & & \left. \qquad \qquad \qquad
{} - \sigma^\lambda p_{\mu_1} 
[ Z(E_+) - Z(E_-) ] p_{\mu_2} Z'(E_-) \right\} .
\label{eq_app_bastin}
\end{eqnarray}
Here the prefactor $c$ is inversely proportional to the size
of the system (a big finite crystal with periodic boundary
conditions), the integration is carried out over the occupied part
of the valence spectrum (for energies $E$ up to the Fermi energy
$E_\mathrm{F}$), and the trace $\overline{\mathrm{Tr}}$ refers
to the Hilbert space of the entire system.
The quantities $\sigma^\lambda$ denote the Pauli spin matrices,
$(\sigma^x , \sigma^y , \sigma^z ) = \bm{\sigma}$, the quantities
$p_\mu$ refer to the momentum operator, $(p_x , p_y , p_z ) = 
\mathbf{p}$, the symbol $Z(E_\pm) = Z(E \pm i 0)$ denotes the
retarded and advanced one-electron propagator (resolvent), and
the prime at $Z(E_\pm)$ denotes energy derivative.
Note that evaluation of Eq.~(\ref{eq_app_bastin}) involves implicitly
averaging over all $\mathbf{k}$ vectors in the whole BZ.
The direction of all magnetic moments (and exchange fields) of the
collinear system is specified by a unit vector $\mathbf{n} = 
(n_x , n_y , n_z )$.
The momentum operator $p_\mu$ is spin independent; the spin
dependence of the propagators $Z(E_\pm)$ can be written as a sum
over the spin channel index $s$ ($s = \pm 1$) as
\begin{equation}
Z(E_\pm) = \sum_s Z_s(E_\pm) \otimes \Pi_s (\mathbf{n}) ,
\qquad
\Pi_s (\mathbf{n}) = 
\frac{1 + s \mathbf{n} \cdot \bm{\sigma}}{2},
\label{eq_app_spindep}
\end{equation}
and similarly for the derivatives $Z'(E_\pm)$.
Here the symbol $a \otimes b$ means an operator involving an
operator $a$ acting only in the orbital space and an operator $b$
acting only in the two-dimensional spin space.
The quantities $Z_s(E_\pm)$ in Eq.~(\ref{eq_app_spindep}) refer thus
to the propagators in the spin channel $s$ ($s = \pm 1$) while the
$\Pi_s (\mathbf{n})$ denotes a projection operator in the spin space
(projecting on the spin channel $s$ with respect to the spin
quantization axis $\mathbf{n}$).
Evaluation of the trace follows the rule
$\overline{\mathrm{Tr}} ( a \otimes b ) = 
\underline{\mathrm{Tr}} (a) \mathrm{tr} (b)$, where the traces
$\underline{\mathrm{Tr}}$ and $\mathrm{tr}$ refer to the orbital
and spin space, respectively.
Using Eq.~(\ref{eq_app_spindep}) in the starting
formula~(\ref{eq_app_bastin}) together with the relation
\begin{equation}
\mathrm{tr} [ \sigma^\lambda 
\Pi_s (\mathbf{n}) \Pi_{s'} (\mathbf{n}) ]
= n_\lambda s \delta_{s s'} 
\label{eq_app_spintrace}
\end{equation}
leads to the final expression for the spin-conductivity tensor
$\sigma^\lambda_{\mu_1 \mu_2}$ as
\begin{eqnarray}
\sigma^\lambda_{\mu_1 \mu_2} & = &
n_\lambda \tilde{\sigma}_{\mu_1 \mu_2} , 
\qquad 
\tilde{\sigma}_{\mu_1 \mu_2} = 
\sum_s s \sigma^{(s)}_{\mu_1 \mu_2} ,
\nonumber\\
\sigma^{(s)}_{\mu_1 \mu_2} & = &
-2c \int^{E_\mathrm{F}}_{-\infty} dE \underline{\mathrm{Tr}}
\left\{ p_{\mu_1} Z'_s(E_+) p_{\mu_2} [ Z_s(E_+) - Z_s(E_-) ] \right.
\nonumber\\
 & & \left. \qquad \qquad \qquad
{} - p_{\mu_1} [ Z_s(E_+) - Z_s(E_-) ] p_{\mu_2} Z'_s(E_-) \right\} .
\label{eq_app_spincond}
\end{eqnarray}
This result proves the reduction of the tensor 
$\sigma^\lambda_{\mu_1 \mu_2}$ of rank three to a tensor
$\tilde{\sigma}_{\mu_1 \mu_2}$ of rank two; the latter equals
the difference of tensors $\sigma^{(s)}_{\mu_1 \mu_2}$ for the
majority (spin-up, $s=1$) and minority (spin-down, $s=-1$) channels.
The tensor $\sigma^{(s)}_{\mu_1 \mu_2}$ coincides with the electrical
conductivity tensor in channel $s$, which can be expressed by the
Kubo-Greenwood formula \cite{r_1958_dag} in terms of the
spin-resolved propagators only at the Fermi energy.
This yields:
\begin{equation}
\sigma^{(s)}_{\mu_1 \mu_2} = c \underline{\mathrm{Tr}}
[ p_{\mu_1} \Gamma_s p_{\mu_2} \Gamma_s ] ,
\qquad
\Gamma_s = i [ Z_s(E_{\mathrm{F},+}) - Z_s(E_{\mathrm{F},-}) ] .
\label{eq_app_kgs0}
\end{equation}
Using the pseudoscalar spin operator $\sigma$, the notation
of Section~\ref{ss_pes}, and an operator $\Gamma$, diagonal in the
spin-channel index and defined by its spin-resolved blocks
$\Gamma_s$, the reduced spin-conductivity tensor
$\tilde{\sigma}_{\mu_1 \mu_2}$ can be rewritten as
\begin{equation}
\tilde{\sigma}_{\mu_1 \mu_2} = c \overline{\mathrm{Tr}}
( \sigma p_{\mu_1} \Gamma p_{\mu_2} \Gamma ) .
\label{eq_app_kg}
\end{equation}

The group invariance of the system means that the Hamiltonian
commutes with a unitary operator $D$ representing the combination
of a rotation $\alpha$, translation $\mathbf{t}$, and spin-channel
interchange $\eta$, see Eq.~(\ref{eq_mpgdef}).
The latter operator is defined by its action on all basic kets as
$D | \mathbf{r}s \rangle = | \mathbf{r}'s' \rangle$, where
$\mathbf{r}' = \alpha \mathbf{r} + \mathbf{t}$ and $s' = \eta s$.
As a consequence, one can derive transformations of the operators
in Eq.~(\ref{eq_app_kg}) as
\begin{equation}
D^{-1} \sigma D = \eta \sigma ,
\qquad 
D^{-1} \Gamma D = \Gamma ,
\qquad
D^{-1} p_\mu D = \sum_\nu \alpha_{\mu \nu} p_\nu ,
\label{eq_app_trpaux}
\end{equation}
in complete analogy with Eq.~(\ref{eq_trpall}).
This leads to a condition for the tensor
$\tilde{\sigma}_{\mu_1 \mu_2}$:
\begin{eqnarray}
\tilde{\sigma}_{\mu_1 \mu_2}
 & = & c \overline{\mathrm{Tr}} ( D \eta \sigma D^{-1} 
 p_{\mu_1} D \Gamma D^{-1} p_{\mu_2} D \Gamma D^{-1} )
\nonumber\\
 & = & \eta \sum_{\nu_1 \nu_2}
 \alpha_{\mu_1 \nu_1} \alpha_{\mu_2 \nu_2} c
\overline{\mathrm{Tr}} ( \sigma p_{\nu_1} \Gamma p_{\nu_2} \Gamma )
 = \eta \sum_{\nu_1 \nu_2} \alpha_{\mu_1 \nu_1}
 \alpha_{\mu_2 \nu_2} \tilde{\sigma}_{\nu_1 \nu_2} ,
\label{eq_app_csct}
\end{eqnarray}
which has the same form as Eq.~(\ref{eq_scn2}) for the tensor
$T^{(2)}_{\mu_1 \mu_2}$.
The derived condition (\ref{eq_app_csct}) does not contain explicitly
the translation vector $\mathbf{t}$, so that it holds for all
elements $( \alpha , \eta )$ of the magnetic point group 
$\mathcal{P}_\mathrm{M}$.
This proves that the shapes of both symmetric tensors
$\tilde{\sigma}_{\mu_1 \mu_2}$ and $T^{(2)}_{\mu_1 \mu_2}$ are
identical.

\section{Detailed results of symmetry analysis\label{app_drsa}}

In this part, we list more details on the resulting nonvanishing
tensors $T^{(N)}_{\mu_1 \mu_2 \dots \mu_N}$, sketched briefly in
Table~\ref{t_cap} for all ten nontrivial magnetic Laue groups.
For groups possessing only one rotation axis of the maximal order,
this axis coincides with $z$ axis; further information on the
orientation of the symmetry elements with respect to the coordinate
system is given below for each particular group.
In listing the independent nonzero tensor components, relations
reflecting the full symmetry of $T^{(N)}_{\mu_1 \mu_2 \dots \mu_N}$,
such as, e.g., $T^{(2)}_{xy} = T^{(2)}_{yx}$, are not mentioned
explicitly.
We also give the leading term of the Taylor expansion of the
function $F(\mathbf{k})$, Eq.~(\ref{eq_tayfk}); the symbols
$c_1$ and $c_2$ below denote two arbitrary constants.

For the group m$'$m$'$m, we chose the unprimed reflection on 
$x - y$ plane and the primed reflections on $x - z$ and $y - z$
planes. 
We get $N = 2$ and a single component $T^{(2)}_{xy}$.
This yields:
\begin{equation}
F(\mathbf{k}) \sim k_x k_y .
\label{eq_app_case_mpmpm}
\end{equation}
For the group 2$'$/m$'$, we  get $N = 2$ and two components,
$T^{(2)}_{xz}$ and $T^{(2)}_{yz}$.
This yields:
\begin{equation}
F(\mathbf{k}) = c_1 k_x k_z + c_2 k_y k_z .
\label{eq_app_case_2psmp}
\end{equation}
For the group 4$'$/m, we get $N = 2$ and two components,
$T^{(2)}_{xx} = - T^{(2)}_{yy}$ and $T^{(2)}_{xy}$.
This yields:
\begin{equation}
F(\mathbf{k}) = c_1 ( k^2_x - k^2_y ) + c_2 k_x k_y .
\label{eq_app_case_4psm}
\end{equation}
For the group 4$'$/mm$'$m, the primed reflection was on (110) plane.
We get $N = 2$ and a single component
$T^{(2)}_{xx} = - T^{(2)}_{yy}$.
This yields:
\begin{equation}
F(\mathbf{k}) \sim k^2_x - k^2_y .
\label{eq_app_case_4psmmpm}
\end{equation}
For the group $\bar{3}$m$'$, the primed reflection was on $y - z$
plane.
We get $N = 4$ and a single component
$T^{(4)}_{xxxz} = - T^{(4)}_{xyyz}$.
This yields:
\begin{equation}
F(\mathbf{k}) \sim k_x k_z ( k^2_x - 3 k^2_y ) .
\label{eq_app_case_b3mp}
\end{equation}
For the group 4/mm$'$m$'$, one of the primed reflections was on
$y - z$ plane. 
We get $N = 4$ and a single component
$T^{(4)}_{xxxy} = - T^{(4)}_{xyyy}$.
This yields:
\begin{equation}
F(\mathbf{k}) \sim k_x k_y ( k^2_x - k^2_y ) .
\label{eq_app_case_4smmpmp}
\end{equation}
For the group 6$'$/m$'$, we get $N = 4$ and two components,
$T^{(4)}_{xxxz} = - T^{(4)}_{xyyz}$ and
$T^{(4)}_{xxyz} = - T^{(4)}_{yyyz}$.
This yields:
\begin{equation}
F(\mathbf{k}) = c_1 k_x k_z ( k^2_x - 3 k^2_y ) 
              + c_2 k_y k_z ( k^2_y - 3 k^2_x ) .
\label{eq_app_case_6psmp}
\end{equation}
For the group 6$'$/m$'$m$'$m, the unprimed reflection was on 
$y - z$ plane. 
We get $N = 4$ and a single component
$T^{(4)}_{xxyz} = - T^{(4)}_{yyyz}$.
This yields:
\begin{equation}
F(\mathbf{k}) \sim k_y k_z ( 3 k^2_x - k^2_y ) .
\label{eq_app_case_6psmpmpm}
\end{equation}
For the group 6/mm$'$m$'$, the primed reflections were on $x - z$
and $y - z$ planes. 
We get $N = 6$ and a single component
$T^{(6)}_{xxxxxy} = - T^{(6)}_{xxxyyy} = T^{(6)}_{xyyyyy}$.
This yields:
\begin{equation}
F(\mathbf{k}) \sim k_x k_y ( 3 k^2_x - k^2_y ) ( 3 k^2_y - k^2_x ) .
\label{eq_app_case_6smmpmp}
\end{equation}
For the group m$\bar{3}$m$'$,
the threefold rotation axes were chosen along (111), (11$\bar{1}$),
(1$\bar{1}$1), and (1$\bar{1}\bar{1}$) directions.
We get $N = 6$ and a single component
$T^{(6)}_{xxxxyy} = - T^{(6)}_{xxxxzz} = - T^{(6)}_{xxyyyy} =
 T^{(6)}_{xxzzzz} =   T^{(6)}_{yyyyzz} = - T^{(6)}_{yyzzzz}$.
This yields:
\begin{equation}
F(\mathbf{k}) \sim 
( k^2_x - k^2_y ) ( k^2_y - k^2_z ) ( k^2_z - k^2_x ) .
\label{eq_app_case_mb3mp}
\end{equation}

The obtained functions $F(\mathbf{k})$ for the individual
magnetic point groups can be compared with their eigenvalue-based
counterparts.
These functions for the groups 
m$'$m$'$m (\ref{eq_app_case_mpmpm}), 
$\bar{3}$m$'$ (\ref{eq_app_case_b3mp}),
4/mm$'$m$'$ (\ref{eq_app_case_4smmpmp}), 
6/mm$'$m$'$ (\ref{eq_app_case_6smmpmp}),
and m$\bar{3}$m$'$ (\ref{eq_app_case_mb3mp})
are identical to those of Ref.~\onlinecite{r_2021_ssj}.
In two other cases, differences are encountered which however can
easily be removed by rotations of the coordinate systems:
for the group 4$'$/mm$'$m (\ref{eq_app_case_4psmmpm})
a rotation by $\pi / 4$ around $z$ axis is needed,
while for the group 6$'$/m$'$m$'$m (\ref{eq_app_case_6psmpmpm})
a rotation by $\pi / 2$ around $z$ axis is needed
(these rotations correspond to an interchange of the secondary
and tertiary symmetry directions for both groups).
In the remaining three cases, i.e., for the groups
2$'$/m$'$ (\ref{eq_app_case_2psmp}),
4$'$/m (\ref{eq_app_case_4psm}), and
6$'$/m$'$ (\ref{eq_app_case_6psmp}),
the derived functions $F(\mathbf{k})$ contain two terms, whereby
only one of them coincides with the corresponding single-term
expression of Ref.~\onlinecite{r_2021_ssj}.
This can be ascribed to the fact that all elements of these point
groups are insensitive to the choice of a direction of $x$ (and $y$)
axis, whereas this ambiguity is always missing in a model
calculation using a particular lattice, which leads to a 
suppression of one of both terms.

\section{(Supplemental Material) Relation between standard
          and modified magnetic point groups}

In this part, we consider a collinear magnet with its effective
one-electron Hamiltonian characterized by a spin-averaged potential
$\bar{V}(\mathbf{r})$ and an exchange field $B(\mathbf{r})$, defined
in terms of the spin-resolved potentials
$V_s(\mathbf{r})$ ($s = \pm 1$) as
\begin{equation}
\bar{V}(\mathbf{r}) = [ V_+(\mathbf{r}) + V_-(\mathbf{r}) ]/2 ,
\qquad
 B(\mathbf{r}) = [ V_+(\mathbf{r}) - V_-(\mathbf{r}) ]/2 ,
\label{eq_sm1_vb}
\end{equation}
see Section II A of the main article.
We assume that both quantities exhibit full three-dimensional
translation invariance and that the exchange field is not
identically zero.
We prove several general relations between the two magnetic
point groups of the system:
the standard one, $\mathcal{P}^\mathrm{st}_\mathrm{M}$, relevant
for the exchange field as a vector quantity, and the modified one,
$\mathcal{P}^\mathrm{mod}_\mathrm{M}$, relevant for the exchange
field as a scalar quantity.

Both groups consist of elements $(\alpha , \eta)$ where $\alpha$
is a real $3\times 3$ orthogonal matrix and $\eta \in \{ 1, -1 \}$ 
indicating the absence (for $\eta = 1$) or presence (for $\eta = -1$)
of the operation of antisymmetry in the group element.
The unit element of both groups is denoted as $(1,1)$, whereas
$(1,-1)$ refers to an element describing the pure operation of
antisymmetry.

The standard magnetic point group
$\mathcal{P}^\mathrm{st}_\mathrm{M}$ of the considered collinear
magnet depends not only on the functions $\bar{V}(\mathbf{r})$
and $B(\mathbf{r})$, but also on the exchange field direction,
which we specify by a unit vector $\mathbf{n}$.
The $\mathcal{P}^\mathrm{st}_\mathrm{M}$ contains then all elements
$(\alpha , \eta)$ such that a translation vector $\mathbf{t}$
[dependent on $(\alpha , \eta)$] exists, for which identities
\begin{equation}
\bar{V}(\mathbf{r}) = \bar{V}(\alpha \mathbf{r} + \mathbf{t}) ,
\qquad 
B(\mathbf{r}) \mathbf{n} = \eta | \alpha |
B(\alpha \mathbf{r} + \mathbf{t}) \alpha^{-1} \mathbf{n} 
\label{eq_sm1_pmst}
\end{equation}
are valid for all $\mathbf{r}$.
Here the symbol $| \alpha |$ denotes the determinant of the
matrix $\alpha$ ($| \alpha | = 1$ for proper rotations,
$| \alpha | = -1$ for improper rotations).
The identity for the exchange field corresponds obviously to its
particular vector nature: the rotation of the field direction is
reflected by the vector $\alpha^{-1} \mathbf{n}$, while the prefactor
$| \alpha |$ reflects the axiality of the exchange field (axial
vectors do not change their sign upon space inversion) and the
prefactor $\eta$ reflects the sign change of the exchange field due
to the operation of antisymmetry (the latter sign change is
equivalent physically to the sign change of a magnetic field due to
time reversal).

The modified magnetic point group
$\mathcal{P}^\mathrm{mod}_\mathrm{M}$ comprises all elements
$(\alpha , \eta)$ such that a translation vector $\mathbf{t}$
[dependent on $(\alpha , \eta)$] exists, for which identities
\begin{equation}
\bar{V}(\mathbf{r}) = \bar{V}(\alpha \mathbf{r} + \mathbf{t}) ,
\qquad
B(\mathbf{r}) = \eta B(\alpha \mathbf{r} + \mathbf{t}) 
\label{eq_sm1_pmmod}
\end{equation}
are valid for all $\mathbf{r}$.
This definition of $\mathcal{P}^\mathrm{mod}_\mathrm{M}$ is
equivalent to that of $\mathcal{P}_\mathrm{M}$ given in the text
around Eqs.\ (4) and (5) in Section II A of the main article.
The identity for the exchange field corresponds obviously to its
specific scalar nature: the prefactor $\eta$ reflects the sign
change of the exchange field due to the operation of antisymmetry.

In addition to both magnetic point groups, we introduce their
parent crystallographic point groups denoted as
$\mathcal{P}^\mathrm{st}$ and $\mathcal{P}^\mathrm{mod}$.
The group $\mathcal{P}^\mathrm{st}$ comprises thus all rotations
$\alpha$ such that $(\alpha , \eta) \in 
\mathcal{P}^\mathrm{st}_\mathrm{M}$ for some $\eta$.
Similarly, the group $\mathcal{P}^\mathrm{mod}$ contains all
rotations $\alpha$ such that $(\alpha , \eta) \in
\mathcal{P}^\mathrm{mod}_\mathrm{M}$ for some $\eta$.
With all these definitions, we can prove following three theorems.

\textbf{Theorem 1.}
The standard parent group $\mathcal{P}^\mathrm{st}$ is a subgroup
of the modified parent group $\mathcal{P}^\mathrm{mod}$.

\textit{Proof.}
Let $\alpha \in \mathcal{P}^\mathrm{st}$, so that an 
$\eta_1 \in \{ 1 , -1 \}$ exists such that $(\alpha , \eta_1 ) \in
\mathcal{P}^\mathrm{st}_\mathrm{M}$.
This means according to Eq.~(\ref{eq_sm1_pmst}) that a translation
$\mathbf{t}$ exists, for which identities
\begin{equation}
\bar{V}(\mathbf{r}) = \bar{V}(\alpha \mathbf{r} + \mathbf{t}) ,
\qquad
B(\mathbf{r}) \mathbf{n} = \eta_1 | \alpha |
B(\alpha \mathbf{r} + \mathbf{t}) \alpha^{-1} \mathbf{n} 
\label{eq_sm1_proof1a}
\end{equation}
hold for all $\mathbf{r}$.
Since the function $B(\mathbf{r})$ is not identically zero,
the unit vectors $\mathbf{n}$ and $\alpha^{-1} \mathbf{n}$ 
must be parallel mutually, so that $\alpha^{-1} \mathbf{n} = 
\varepsilon \mathbf{n}$ for an $\varepsilon \in \{ 1 , -1 \}$.
Since $\mathbf{n}$ is a nonzero vector, we get from
Eq.~(\ref{eq_sm1_proof1a}) for all $\mathbf{r}$ the identities
\begin{equation}
\bar{V}(\mathbf{r}) = \bar{V}(\alpha \mathbf{r} + \mathbf{t}) ,
\qquad
B(\mathbf{r}) = \eta_1 | \alpha | \varepsilon
B(\alpha \mathbf{r} + \mathbf{t}) 
 = \eta B(\alpha \mathbf{r} + \mathbf{t}) ,
\label{eq_sm1_proof1b}
\end{equation}
where $\eta = \eta_1 | \alpha | \varepsilon \in \{ 1 , -1 \}$.
This means according to Eq.~(\ref{eq_sm1_pmmod}) that
$(\alpha , \eta) \in \mathcal{P}^\mathrm{mod}_\mathrm{M}$. 
This yields thus $\alpha \in \mathcal{P}^\mathrm{mod}$, which
completes the proof.

\textit{Consequence.}
The group-subgroup relation provided by Theorem~1 implies that
the order $| \mathcal{P}^\mathrm{st} |$ of the standard parent
group divides the order $| \mathcal{P}^\mathrm{mod} |$ of the
modified parent group.

\textbf{Theorem 2.}
The standard magnetic point group
$\mathcal{P}^\mathrm{st}_\mathrm{M}$ belongs to category (a)
if and only if the modified magnetic point group
$\mathcal{P}^\mathrm{mod}_\mathrm{M}$ belongs to category (a).
 
\textit{Proof.}
(i) Let us assume first, that the modified magnetic point group
$\mathcal{P}^\mathrm{mod}_\mathrm{M}$ belongs to category (a),
so that $(1,-1) \in \mathcal{P}^\mathrm{mod}_\mathrm{M}$.
Then, following Eq.~(\ref{eq_sm1_pmmod}), a translation $\mathbf{t}$
exists, for which identities
\begin{equation}
\bar{V}(\mathbf{r}) = \bar{V}(\mathbf{r} + \mathbf{t}) ,
\qquad 
B(\mathbf{r}) = - B(\mathbf{r} + \mathbf{t}) 
\label{eq_sm1_proof2a}
\end{equation}
hold for all $\mathbf{r}$.
This yields identities
\begin{equation}
\bar{V}(\mathbf{r}) = \bar{V}(\mathbf{r} + \mathbf{t}) ,
\qquad 
B(\mathbf{r}) \mathbf{n} =
 - B(\mathbf{r} + \mathbf{t}) \mathbf{n}  
\label{eq_sm1_proof2b}
\end{equation}
valid for all $\mathbf{r}$.
This means according to Eq.~(\ref{eq_sm1_pmst}) that
$(1, -1) \in \mathcal{P}^\mathrm{st}_\mathrm{M}$. 
Hence, the standard magnetic point group
$\mathcal{P}^\mathrm{st}_\mathrm{M}$ belongs to category (a).

(ii) Let us assume now, that the standard magnetic point group
$\mathcal{P}^\mathrm{st}_\mathrm{M}$ belongs to category (a),
so that $(1,-1) \in \mathcal{P}^\mathrm{st}_\mathrm{M}$.
Then, following Eq.~(\ref{eq_sm1_pmst}), a translation $\mathbf{t}$
exists, for which identities (\ref{eq_sm1_proof2b}) hold for
all $\mathbf{r}$.
Since the vector $\mathbf{n}$ is nonzero, this implies the validity
of identities (\ref{eq_sm1_proof2a}) for all $\mathbf{r}$.
This means according to Eq.~(\ref{eq_sm1_pmmod}) that
$(1, -1) \in \mathcal{P}^\mathrm{mod}_\mathrm{M}$. 
Hence, the modified magnetic point group
$\mathcal{P}^\mathrm{mod}_\mathrm{M}$ belongs to category (a).
This completes the proof.

\textbf{Theorem 3.}
The order $| \mathcal{P}^\mathrm{st}_\mathrm{M} |$ of the standard
magnetic point group divides the order
$| \mathcal{P}^\mathrm{mod}_\mathrm{M} |$ of the modified
magnetic point group.
 
\textit{Proof.}
This theorem follows from both previous ones.
According to Theorem~2, we can confine ourselves to two separate
cases.
In the first case, both $\mathcal{P}^\mathrm{st}_\mathrm{M}$ and
$\mathcal{P}^\mathrm{mod}_\mathrm{M}$ belong to category (a).
Their orders are then related to those of their parent counterparts
as
\begin{equation}
| \mathcal{P}^\mathrm{st}_\mathrm{M} | = 
2 | \mathcal{P}^\mathrm{st} | ,
\qquad
| \mathcal{P}^\mathrm{mod}_\mathrm{M} | = 
2 | \mathcal{P}^\mathrm{mod} | . 
\label{eq_sm1_proof3a}
\end{equation}
In the second case, none of the groups
$\mathcal{P}^\mathrm{st}_\mathrm{M}$ and
$\mathcal{P}^\mathrm{mod}_\mathrm{M}$ belongs to category (a).
Their orders are then related to those of their parent counterparts
as
\begin{equation}
| \mathcal{P}^\mathrm{st}_\mathrm{M} | = 
| \mathcal{P}^\mathrm{st} | ,
\qquad
| \mathcal{P}^\mathrm{mod}_\mathrm{M} | = 
| \mathcal{P}^\mathrm{mod} | . 
\label{eq_sm1_proof3b}
\end{equation}
Since $| \mathcal{P}^\mathrm{st} |$ always divides
$| \mathcal{P}^\mathrm{mod} |$ (according to the
consequence of Theorem~1), we get in both separate cases that
$| \mathcal{P}^\mathrm{st}_\mathrm{M} |$ divides
$| \mathcal{P}^\mathrm{mod}_\mathrm{M} |$.
This completes the proof.

Two remarks are now in order.
First, the proved relations hold irrespective of the particular
direction $\mathbf{n}$ of the exchange field, whereas the standard
groups $\mathcal{P}^\mathrm{st}_\mathrm{M}$ and 
$\mathcal{P}^\mathrm{st}$ depend on this direction.
Second, the examples of both kinds of magnetic point groups for
selected systems, listed in Table I of the main article, are fully
consistent with the proved theorems.

\section{(Supplemental Material) Relation between spin groups
         and modified magnetic groups}

In this part, we outline relations between the spin groups and
the modified magnetic groups for a collinear nonrelativistic magnet
described by the one-electron Hamiltonian given in Section II A of
the main article.

\subsection{Definitions and auxiliary relations}

The position vector is denoted by $\mathbf{r}$ and
translation vectors in the real space are denoted by $\mathbf{t}$
(and also by $\mathbf{t}_1$, $\mathbf{t}_2$, \dots).  
The symbol $s$ (and also $s'$) denotes the spin-channel index,
$s \in \{1, -1\}$, and $| s \rangle$ denotes the corresponding basis
vector in the two-dimensional spin space for a particle with
spin 1/2.
The symbol $\alpha$ (and also $\alpha_1$, $\alpha_2$, \dots)
denotes a rotation in the real space, i.e., a real orthogonal
$3 \times 3$ matrix, $\alpha \in O(3)$.
The symbol $\beta$ (and also $\beta_1$, $\beta_2$, \dots) denotes
a unitary or antiunitary operator in the two-dimensional spin space,
$\beta \in \bar{U}(2)$, where $\bar{U}(2)$ denotes the group of all
such operators.
The symbol $\eta$ (and also $\eta_1$, $\eta_2$, \dots) denotes a
discrete variable that acquires two values, $\eta \in \{ 1 , -1 \}$.

The symbol $V$ denotes an operator (spin-dependent local one-electron
potential) acting in the full vector space spanned by the basic kets
$| \mathbf{r}s \rangle$; its action is defined by
\begin{equation}
V | \mathbf{r}s \rangle = V_s (\mathbf{r}) | \mathbf{r}s \rangle ,
\label{eq_sm2_vdef}
\end {equation}
where $V_s (\mathbf{r})$, $s = \pm 1$, are two different real
functions [so that $V_+ (\mathbf{r}) \neq V_- (\mathbf{r})$ for some
$\mathbf{r}$].
In the two-dimensional spin space, we introduce two unitary
operators, $\sigma$ (spin operator) and $\omega$ (operator of
spin-channel interchange), defined by
\begin{equation}
\sigma | s \rangle = s | s \rangle ,
\qquad
\omega | s \rangle = | \! - \! s \rangle ,
\label{eq_sm2_sodef}
\end {equation}
valid for both values of $s$.
Their properties include relations
\begin{equation}
\sigma^2 = \omega^2 = 1 ,
\qquad
\sigma \omega = - \omega \sigma .
\label{eq_sm2_soprop}
\end {equation}

Besides the group $\bar{U}(2)$, we introduce its two subsets
$\bar{U}(2, \eta)$, $\eta = \pm 1$; the subset $\bar{U}(2, \eta)$
comprises all operators $\beta \in \bar{U}(2)$ satisfying
\begin{equation}
\sigma \beta = \eta \beta \sigma .
\label{eq_sm2_bu2eta}
\end {equation}
In other words, the $\bar{U}(2, 1)$ contains all operators $\beta$
commuting with $\sigma$, while the $\bar{U}(2, -1)$ contains all
operators $\beta$ anticommuting with $\sigma$.
The subset $\bar{U}(2, 1)$ is a subgroup of $\bar{U}(2)$.
It can easily be proved in terms of matrix elements
$\beta_{s's} = \langle s' | \beta | s \rangle$ that an operator
$\beta \in \bar{U}(2)$ belongs to $\bar{U}(2, 1)$ if and only if it
is diagonal in the spin indices (so that $\langle -s | \beta | s
\rangle = 0$ for both $s$).
Similarly, an operator $\beta \in \bar{U}(2)$ belongs to
$\bar{U}(2, -1)$ if and only if it is purely nondiagonal in the spin
indices (so that $\langle s | \beta | s \rangle = 0$ for both $s$).
Alternatively expressed, $\beta \in \bar{U}(2, \eta)$ means that
$\beta | s \rangle$ is proportional to $| \eta s \rangle$ for both
basis vectors $| s \rangle$.
These facts lead to the following lemma.

\textbf{Lemma 1.}
Any operator $\beta \in \bar{U}(2)$ either belongs to
$\bar{U}(2, \eta)$ for some $\eta \in \{ 1, -1 \}$, or it has all of
its matrix elements nonzero ($\beta_{s's} \neq 0$ for all pairs of
$s', s$).

Moreover, it can be shown that for $\beta \in \bar{U}(2, \eta)$, one
obtains $\beta^{-1} \in \bar{U}(2, \eta)$, and for
$\beta_1 \in \bar{U}(2, \eta_1)$ and
$\beta_2 \in \bar{U}(2, \eta_2)$, one obtains
$\beta_1 \beta_2 \in \bar{U}(2, \eta_1 \eta_2)$.
Note that $\sigma \in \bar{U}(2, 1)$ and $\omega \in \bar{U}(2, -1)$.

We also introduce a mapping $\eta \mapsto \gamma(\eta)$ of the
discrete variable $\eta$ into the group $\bar{U}(2)$:
\begin{equation}
\gamma(1) = 1 ,
\qquad
\gamma(-1) = \omega .
\label{eq_sm2_gammadef}
\end{equation}
This mapping satisfies 
\begin{equation}
\gamma(\eta_1 \eta_2) = \gamma(\eta_1) \gamma(\eta_2) ,
\label{eq_sm2_gamma12}
\end{equation}
and $\gamma^2(\eta) = 1$ as well as
$\gamma(\eta) \in \bar{U}(2,\eta)$ for both values of $\eta$
($\eta = \pm 1$).

\subsection{Space groups}

For the definition of the spin space group $\mathcal{G}_\mathrm{S}$
of a collinear nonrelativistic magnet, let us introduce first a
bigger group $\mathcal{G}^\infty_\mathrm{S}$ containing all
elements of the type $(\alpha, \mathbf{t} | \beta )$ with a group
multiplication rule
\begin{equation}
(\alpha_1, \mathbf{t}_1 | \beta_1 )
(\alpha_2, \mathbf{t}_2 | \beta_2 ) =
(\alpha_1 \alpha_2, \alpha_1 \mathbf{t}_2 + \mathbf{t}_1 |
 \beta_1 \beta_2 ) .
\label{eq_sm2_gmrss}
\end{equation}
For each element $(\alpha, \mathbf{t} | \beta ) \in 
\mathcal{G}^\infty_\mathrm{S}$, we define an operator
$\Gamma(\alpha, \mathbf{t} | \beta )$ by its action on the
vectors $| \mathbf{r} \rangle \otimes | \chi \rangle$, where
$| \chi \rangle$ denotes a vector of the two-dimensional spin space,
or equivalently on the basic kets $| \mathbf{r}s \rangle =
| \mathbf{r} \rangle \otimes | s \rangle$ as follows:
\begin{equation}
\Gamma(\alpha, \mathbf{t} | \beta ) | \mathbf{r} \rangle
\otimes | \chi \rangle = 
| \alpha \mathbf{r} + \mathbf{t} \rangle 
\otimes \beta | \chi \rangle ,
\qquad
\Gamma(\alpha, \mathbf{t} | \beta ) | \mathbf{r}s \rangle = 
\sum_{s'} | \alpha \mathbf{r} + \mathbf{t} , s' \rangle \beta_{s's} .
\label{eq_sm2_biggamma}
\end{equation}
This operator is unitary or antiunitary for a unitary or antiunitary
$\beta \in \bar{U}(2)$, respectively. 
It can be shown that Eq.~(\ref{eq_sm2_biggamma}) defines a
representation of the big group $\mathcal{G}^\infty_\mathrm{S}$;
in particular, an operator counterpart of Eq.~(\ref{eq_sm2_gmrss}),
\begin{equation}
\Gamma (\alpha_1, \mathbf{t}_1 | \beta_1 )
\Gamma (\alpha_2, \mathbf{t}_2 | \beta_2 ) =
 \Gamma (\alpha_1 \alpha_2, \alpha_1 \mathbf{t}_2 + \mathbf{t}_1 |
 \beta_1 \beta_2 ) , 
\label{eq_sm2_gammarep}
\end{equation}
is valid.
The spin space group $\mathcal{G}_\mathrm{S}$ of a particular system
is now defined as a subgroup of $\mathcal{G}^\infty_\mathrm{S}$
comprising all its elements $(\alpha, \mathbf{t} | \beta )$
satisfying
\begin{equation}
\Gamma(\alpha, \mathbf{t} | \beta ) V =
V \Gamma(\alpha, \mathbf{t} | \beta ).
\label{eq_sm2_gssdef}
\end{equation}
Since the kinetic term in the one-electron Hamiltonian is invariant
to all elements of the group $\mathcal{G}^\infty_\mathrm{S}$, the
last relation is equivalent to the invariance of the total
Hamiltonian with respect to the element of the spin space group
$\mathcal{G}_\mathrm{S}$.

For the definition of the modified magnetic space group
$\mathcal{G}_\mathrm{M}$ of the same system, let us introduce
first a bigger group $\mathcal{G}^\infty_\mathrm{M}$ containing all
elements of the type $(\alpha, \mathbf{t} , \eta )$ with a group
multiplication rule
\begin{equation}
(\alpha_1, \mathbf{t}_1 , \eta_1 )
(\alpha_2, \mathbf{t}_2 , \eta_2 ) =
(\alpha_1 \alpha_2, \alpha_1 \mathbf{t}_2 + \mathbf{t}_1 ,
 \eta_1 \eta_2 ) .
\label{eq_sm2_gmrms}
\end{equation}
The modified magnetic space group $\mathcal{G}_\mathrm{M}$ of the
system is now defined as a subgroup of
$\mathcal{G}^\infty_\mathrm{M}$
comprising all its elements $(\alpha, \mathbf{t} , \eta )$ such,
that  
\begin{equation}
V_s(\mathbf{r}) = V_{\eta s}(\alpha \mathbf{r} + \mathbf{t})
\label{eq_sm2_gmsdef}
\end{equation}
holds for all vectors $\mathbf{r}$ and both values of $s$
($s = \pm 1$).
The adopted assumption of different functions $V_+(\mathbf{r})$ and
$V_-(\mathbf{r})$ is thus equivalent to the assumption that the
pure operation of antisymmetry is not an element of the modified
magnetic space group, $(1, \textbf{0} , -1) \notin 
\mathcal{G}_\mathrm{M}$.
Note that the condition~(\ref{eq_sm2_gmsdef}) coincides with Eq.~(5)
of the main article.
The mutual relation between both introduced space groups is
described by the following theorem.

\textbf{Theorem 4.}
An element $(\alpha, \mathbf{t} | \beta ) \in 
\mathcal{G}^\infty_\mathrm{S}$ belongs to the spin space group
$\mathcal{G}_\mathrm{S}$ if and only if the operator $\beta \in
\bar{U}(2)$ belongs to the subset $\bar{U}(2, \eta)$ for some $\eta
\in \{1, -1\}$ and the element $(\alpha, \mathbf{t} , \eta ) \in
\mathcal{G}^\infty_\mathrm{M}$ belongs to the modified magnetic
space group $\mathcal{G}_\mathrm{M}$.

\textit{Proof.} 
First, we assume that $(\alpha, \mathbf{t} | \beta ) \in
\mathcal{G}_\mathrm{S}$.
Let us take an arbitrary basic ket $| \mathbf{r}s \rangle$ and let us
act by operators on both sides of Eq.~(\ref{eq_sm2_gssdef}) on it.
We get:
\begin{equation}
\Gamma(\alpha, \mathbf{t} | \beta ) V | \mathbf{r}s \rangle
= V_s (\mathbf{r}) \sum_{s'}
| \alpha \mathbf{r} + \mathbf{t} , s' \rangle \beta_{s's} ,
\label{eq_sm2_proof4a}
\end{equation}
and
\begin{equation}
V \Gamma(\alpha, \mathbf{t} | \beta ) | \mathbf{r}s \rangle
= \sum_{s'} V_{s'} (\alpha \mathbf{r} + \mathbf{t})
| \alpha \mathbf{r} + \mathbf{t} , s' \rangle \beta_{s's} .
\label{eq_sm2_proof4b}
\end{equation}
Since both results must be the same (for all kets
$| \mathbf{r}s \rangle$), we get the relation
\begin{equation}
V_s (\mathbf{r}) \beta_{s's} = 
V_{s'} (\alpha \mathbf{r} + \mathbf{t}) \beta_{s's}
\label{eq_sm2_proof4c}
\end{equation}
valid for all $\mathbf{r}$ and for all pairs of $s, s'$ 
($s, s' \in \{1, -1 \}$).
If all matrix elements $\beta_{s's}$ were nonzero, then we would
have $V_s (\mathbf{r}) = V_{s'} (\alpha \mathbf{r} + \mathbf{t})$
for all $\mathbf{r}$ and all pairs of $s, s'$.
This would result in $V_+ (\mathbf{r}) = V_- (\mathbf{r})$ valid
for all $\mathbf{r}$, which contradicts the assumed different
potentials $V_+ (\mathbf{r})$ and $V_- (\mathbf{r})$.
This means according to Lemma~1, that we must have 
$\beta \in \bar{U}(2, \eta )$ for some $\eta \in \{1 , -1 \}$.
Consequently, the only nonzero matrix elements of $\beta$ are
$\langle \eta s | \beta | s \rangle$ for $s \in \{1 , -1 \}$.
The relation~(\ref{eq_sm2_proof4c}) yields then
Eq.~(\ref{eq_sm2_gmsdef}) valid for all $\mathbf{r}$ and both
values of $s$, which means that $(\alpha, \mathbf{t} , \eta ) \in
\mathcal{G}_\mathrm{M}$.
The first part of the theorem is proved.

Second, we assume that $(\alpha, \mathbf{t} , \eta ) \in
\mathcal{G}_\mathrm{M}$ and $\beta \in \bar{U}(2, \eta)$.
Let us take an arbitrary basic ket $| \mathbf{r}s \rangle$ and let us
act by operators on both sides of Eq.~(\ref{eq_sm2_gssdef}) on it
using the fact that $\beta | s \rangle$ yields a vector proportional
to $| \eta s \rangle$.
We get:
\begin{equation}
\Gamma(\alpha, \mathbf{t} | \beta ) V | \mathbf{r}s \rangle
= V_s (\mathbf{r}) 
| \alpha \mathbf{r} + \mathbf{t} , \eta s \rangle 
\beta_{\eta s , s} ,
\label{eq_sm2_proof4e}
\end{equation}
and
\begin{equation}
V \Gamma(\alpha, \mathbf{t} | \beta ) | \mathbf{r}s \rangle
= V_{\eta s} (\alpha \mathbf{r} + \mathbf{t})
| \alpha \mathbf{r} + \mathbf{t} , \eta s \rangle 
\beta_{\eta s , s} .
\label{eq_sm2_proof4f}
\end{equation}
Both results are the same because of Eq.~(\ref{eq_sm2_gmsdef}) valid
for $(\alpha, \mathbf{t} , \eta ) \in \mathcal{G}_\mathrm{M}$.
Since this can be done for an arbitrary basic ket
$| \mathbf{r}s \rangle$, it means that the operator
relation~(\ref{eq_sm2_gssdef}) is satisfied and
$(\alpha, \mathbf{t} | \beta ) \in \mathcal{G}_\mathrm{S}$.
This completes the proof of the second part and of the whole theorem.

\textit{Consequence.}
This theorem shows that elements of the modified magnetic space
group $\mathcal{G}_\mathrm{M}$ bear only reduced information as
compared to those of the spin space group $\mathcal{G}_\mathrm{S}$:
instead of the full operator $\beta$ in the element
$(\alpha, \mathbf{t} | \beta )$, only the information about
its spin conservation ($\eta = 1$) or spin interchange ($\eta = -1$)
is kept in the discrete variable $\eta$ of the element
$(\alpha, \mathbf{t} , \eta )$.

Let us consider further a mapping of the big modified magnetic space
group $\mathcal{G}^\infty_\mathrm{M}$ into the big spin space group 
$\mathcal{G}^\infty_\mathrm{S}$ induced by the mapping $\eta \mapsto
\gamma(\eta)$, Eq.~(\ref{eq_sm2_gammadef}):
\begin{equation}
(\alpha , \mathbf{t} , \eta) \mapsto
(\alpha , \mathbf{t} | \gamma(\eta) ) .
\label{eq_sm2_maps}
\end{equation}
Its properties are summarized by the following theorem.

\textbf{Theorem 5.} \\
(i) The mapping~(\ref{eq_sm2_maps}) yields a one-to-one mapping
of the group $\mathcal{G}_\mathrm{M}$ onto a subset 
$\mathcal{G}^\mathrm{M}_\mathrm{S}$ of the group
$\mathcal{G}_\mathrm{S}$. \\
(ii) The set $\mathcal{G}^\mathrm{M}_\mathrm{S}$ is a subgroup of
$\mathcal{G}_\mathrm{S}$. \\
(iii) The groups $\mathcal{G}_\mathrm{M}$ and
$\mathcal{G}^\mathrm{M}_\mathrm{S}$ are isomorphic.

\textit{Proof.} 
(i) Let $(\alpha, \mathbf{t} , \eta ) \in \mathcal{G}_\mathrm{M}$.
Since $\gamma(\eta) \in \bar{U}(2, \eta)$, we have according to
Theorem~4 that $(\alpha , \mathbf{t} | \gamma(\eta) ) \in
\mathcal{G}_\mathrm{S}$.
This means that $\mathcal{G}^\mathrm{M}_\mathrm{S} \subset
\mathcal{G}_\mathrm{S}$.
The one-to-one feature of the mapping~(\ref{eq_sm2_maps}) follows
from the one-to-one feature of the mapping $\eta \mapsto
\gamma(\eta)$, Eq.~(\ref{eq_sm2_gammadef}), which proves the first
part of the theorem.

(ii) First, the identity element $(1, \mathbf{0} , 1) \in
\mathcal{G}_\mathrm{M}$  is mapped on the identity element
$(1 , \mathbf{0} | 1 ) \in \mathcal{G}_\mathrm{S}$ 
which proves that the set $\mathcal{G}^\mathrm{M}_\mathrm{S}$
contains the identity element.
Second, let $(\alpha_j , \mathbf{t}_j | \beta_j ) \in
\mathcal{G}^\mathrm{M}_\mathrm{S}$ for $j = 1, 2$.
Then $\eta_j$ exists such, that $\beta_j = \gamma(\eta_j)$ and
$(\alpha_j , \mathbf{t}_j , \eta_j ) \in
\mathcal{G}_\mathrm{M}$ for $j = 1, 2$.
Let us define $(\alpha_3 , \mathbf{t}_3 | \beta_3 ) =
(\alpha_1 , \mathbf{t}_1 | \beta_1 )
(\alpha_2 , \mathbf{t}_2 | \beta_2 ) \in \mathcal{G}_\mathrm{S}$ 
with the rule~(\ref{eq_sm2_gmrss}), which yields
$\beta_3 = \beta_1 \beta_2$.
Then $(\alpha_1 , \mathbf{t}_1 , \eta_1 )
(\alpha_2 , \mathbf{t}_2 , \eta_2 ) =
(\alpha_3 , \mathbf{t}_3 , \eta_3 ) \in \mathcal{G}_\mathrm{M}$
with $\eta_3 = \eta_1 \eta_2$ according to Eq.~(\ref{eq_sm2_gmrms}).
Employing the rule~(\ref{eq_sm2_gamma12}) we get 
$\beta_3 = \gamma(\eta_3)$, which means that 
$(\alpha_3 , \mathbf{t}_3 | \beta_3 ) \in 
\mathcal{G}^\mathrm{M}_\mathrm{S}$, so that the product of two
elements of $\mathcal{G}^\mathrm{M}_\mathrm{S}$ also belongs to
$\mathcal{G}^\mathrm{M}_\mathrm{S}$.
Third, let $(\alpha , \mathbf{t} | \beta ) \in
\mathcal{G}^\mathrm{M}_\mathrm{S}$, so that an $\eta$ exists such,
that $\beta = \gamma(\eta)$ and $(\alpha , \mathbf{t} , \eta ) \in
\mathcal{G}_\mathrm{M}$.
Then the inverse element to $(\alpha , \mathbf{t} , \eta )$ also
belongs to $\mathcal{G}_\mathrm{M}$, so that $(\alpha^{-1} ,
- \alpha^{-1} \mathbf{t} , \eta ) \in \mathcal{G}_\mathrm{M}$, and,
consequently, $(\alpha^{-1} , - \alpha^{-1} \mathbf{t} | 
\gamma(\eta) ) \in \mathcal{G}^\mathrm{M}_\mathrm{S}$.
However, we have $\gamma(\eta) = \gamma^{-1}(\eta) = \beta^{-1}$,
so that the element 
$(\alpha^{-1} , - \alpha^{-1} \mathbf{t} | \beta^{-1} )$, which
is the inverse of the element $(\alpha , \mathbf{t} | \beta )$,
belongs to $\mathcal{G}^\mathrm{M}_\mathrm{S}$ as well.
This completes the proof of the second part of the theorem.

(iii) Let $(\alpha_j , \mathbf{t}_j , \eta_j ) \in
\mathcal{G}_\mathrm{M}$ and $(\alpha_j , \mathbf{t}_j | \beta_j )
\in \mathcal{G}^\mathrm{M}_\mathrm{S}$, where 
$\beta_j = \gamma(\eta_j)$ for $j = 1, 2, 3$.
Let $(\alpha_1 , \mathbf{t}_1 , \eta_1 )
(\alpha_2 , \mathbf{t}_2 , \eta_2 ) =
(\alpha_3 , \mathbf{t}_3 , \eta_3 )$
according to Eq.~(\ref{eq_sm2_gmrms}).
Then $\eta_3 = \eta_1 \eta_2$ and, using Eq.~(\ref{eq_sm2_gamma12}),
$\beta_3 = \gamma(\eta_3) = \gamma(\eta_1) \gamma(\eta_2) =
\beta_1 \beta_2$, which means that 
$(\alpha_1 , \mathbf{t}_1 | \beta_1 )
(\alpha_2 , \mathbf{t}_2 | \beta_2 ) =
(\alpha_3 , \mathbf{t}_3 | \beta_3 )$
according to Eq.~(\ref{eq_sm2_gmrss}).
This proves the isomorphism and it completes the proof of the whole
theorem.

Let us further consider elements of the type
$( 1 , \mathbf{0} | \beta )$; Theorem~4 yields
$( 1 , \mathbf{0} | \beta ) \in \mathcal{G}_\mathrm{S}$ for any
$\beta \in \bar{U}(2, 1)$.
This enables us to formulate yet another relation between the
introduced space groups.

\textbf{Theorem 6.}
Each element of the spin space group,
$(\alpha, \mathbf{t} | \beta ) \in \mathcal{G}_\mathrm{S}$, can be
written as a product of two elements: an element
$( 1 , \mathbf{0} | \beta_1 )$, where $\beta_1 \in \bar{U}(2, 1)$,
and an element of the group $\mathcal{G}^\mathrm{M}_\mathrm{S}$
isomorphic with the modified magnetic space group
$\mathcal{G}_\mathrm{M}$, $(\alpha, \mathbf{t} | \beta_2 ) \in
\mathcal{G}^\mathrm{M}_\mathrm{S}$.

\textit{Proof.} 
Let $(\alpha, \mathbf{t} | \beta ) \in \mathcal{G}_\mathrm{S}$,
then (from Theorem~4) an $\eta$ exists such that
$\beta \in \bar{U}(2, \eta)$ and $(\alpha, \mathbf{t} , \eta ) \in
\mathcal{G}_\mathrm{M}$.
Let us define $\beta_2 = \gamma(\eta)$, so that $(\alpha, \mathbf{t}
| \beta_2 ) \in \mathcal{G}^\mathrm{M}_\mathrm{S}$ and
$\beta_2 \in \bar{U}(2, \eta)$.
Let us take further $\beta_1 = \beta \beta_2$ which yields  
$\beta_1 \in \bar{U}(2, 1)$.
Moreover, we get $\beta^2_2 = \gamma^2(\eta) = 1$ and 
$\beta_1 \beta_2 = \beta$, so that
$( 1 , \mathbf{0} | \beta_1 ) (\alpha, \mathbf{t} | \beta_2 ) 
= (\alpha, \mathbf{t} | \beta )$, which completes the proof.

\textit{Consequence.}
This theorem shows that the modified magnetic space group
$\mathcal{G}_\mathrm{M}$ forms a skeleton for the whole spin
space group $\mathcal{G}_\mathrm{S}$, since the 'difference'
between the group $\mathcal{G}_\mathrm{S}$ and its subgroup
$\mathcal{G}^\mathrm{M}_\mathrm{S}$ (isomorphic with
$\mathcal{G}_\mathrm{M}$) is the subgroup of elements of the type
$( 1 , \mathbf{0} | \beta )$, where $\beta$ is an element of the
system-independent group $\bar{U}(2, 1)$.

\subsection{Point groups}

Let us consider now the point groups derived from the corresponding
space groups.
We introduce first a bigger group $\mathcal{P}^\infty_\mathrm{S}$
containing all elements of the type $(\alpha | \beta )$ with a group
multiplication rule
\begin{equation}
(\alpha_1 | \beta_1 ) (\alpha_2 | \beta_2 ) =
(\alpha_1 \alpha_2 | \beta_1 \beta_2 ) .
\label{eq_sm2_gmrsp}
\end{equation}
The spin point group $\mathcal{P}_\mathrm{S}$ of a system is defined
as a subgroup of $\mathcal{P}^\infty_\mathrm{S}$ comprising all its
elements $(\alpha | \beta )$ for which a translation $\mathbf{t}$
exists such, that $(\alpha, \mathbf{t} | \beta ) \in 
\mathcal{G}_\mathrm{S}$.

For the definition of the modified magnetic point group
$\mathcal{P}_\mathrm{M}$ of the same system, we introduce first a
bigger group $\mathcal{P}^\infty_\mathrm{M}$ containing all elements
of the type $(\alpha, \eta )$ with a group multiplication rule
\begin{equation}
(\alpha_1, \eta_1 ) (\alpha_2, \eta_2 ) =
(\alpha_1 \alpha_2, \eta_1 \eta_2 ) .
\label{eq_sm2_gmrmp}
\end{equation}
The modified magnetic point group $\mathcal{P}_\mathrm{M}$ is defined
as a subgroup of $\mathcal{P}^\infty_\mathrm{M}$ comprising all its
elements $(\alpha , \eta )$ for which a translation $\mathbf{t}$
exists such, that $(\alpha, \mathbf{t} , \eta ) \in
\mathcal{G}_\mathrm{M}$.
This definition of $\mathcal{P}_\mathrm{M}$ coincides with that given
in Section II A of the main article.
The mutual relation between both introduced point groups is
described by the following theorem.

\textbf{Theorem 7.}
An element $(\alpha | \beta ) \in \mathcal{P}^\infty_\mathrm{S}$
belongs to the spin point group $\mathcal{P}_\mathrm{S}$ if and only
if the operator $\beta \in \bar{U}(2)$ belongs to the subset
$\bar{U}(2, \eta)$ for some $\eta \in \{1, -1\}$ and the element
$(\alpha, \eta ) \in \mathcal{P}^\infty_\mathrm{M}$ belongs to the
modified magnetic point group $\mathcal{P}_\mathrm{M}$.

\textit{Proof.} 
First, we assume that $(\alpha | \beta ) \in \mathcal{P}_\mathrm{S}$.
Then a translation $\mathbf{t}$ exists such, that
$(\alpha, \mathbf{t} | \beta ) \in \mathcal{G}_\mathrm{S}$.
This means according to Theorem~4 that an $\eta$ exists ($\eta \in
\{ 1 , -1 \}$) such that $\beta \in \bar{U}(2, \eta)$ and 
$(\alpha, \mathbf{t} , \eta ) \in \mathcal{G}_\mathrm{M}$.
This yields $(\alpha, \eta ) \in \mathcal{P}_\mathrm{M}$ which
proves the first part of the theorem.

Second, we assume that an $\eta$ exists ($\eta \in \{ 1 , -1 \}$)
such that $\beta \in \bar{U}(2, \eta)$ and $(\alpha, \eta ) \in
\mathcal{P}_\mathrm{M}$.
Then a translation $\mathbf{t}$ exists such, that 
$(\alpha, \mathbf{t} , \eta ) \in \mathcal{G}_\mathrm{M}$.
This means according to Theorem~4 that 
$(\alpha, \mathbf{t} | \beta ) \in \mathcal{G}_\mathrm{S}$, so that
$(\alpha | \beta ) \in \mathcal{P}_\mathrm{S}$.
This proves the second part of the theorem; the proof of the theorem
is now complete.

\textit{Consequence.}
In analogy to Theorem~4, this theorem describes the reduction of
information between the elements
$(\alpha | \beta ) \in \mathcal{P}_\mathrm{S}$ and
$(\alpha, \eta ) \in \mathcal{P}_\mathrm{M}$.

Let us consider further a mapping of the big modified magnetic point
group $\mathcal{P}^\infty_\mathrm{M}$ into the big spin point group 
$\mathcal{P}^\infty_\mathrm{S}$ induced by the mapping $\eta \mapsto
\gamma(\eta)$, Eq.~(\ref{eq_sm2_gammadef}):
\begin{equation}
(\alpha , \eta) \mapsto (\alpha | \gamma(\eta) ) .
\label{eq_sm2_mapp}
\end{equation}
Its properties are summarized by the following theorem.

\textbf{Theorem 8.} \\
(i) The mapping~(\ref{eq_sm2_mapp}) yields a one-to-one mapping
of the group $\mathcal{P}_\mathrm{M}$ onto a subset 
$\mathcal{P}^\mathrm{M}_\mathrm{S}$ of the group
$\mathcal{P}_\mathrm{S}$. \\
(ii) The set $\mathcal{P}^\mathrm{M}_\mathrm{S}$ is a subgroup of
$\mathcal{P}_\mathrm{S}$. \\
(iii) The groups $\mathcal{P}_\mathrm{M}$ and
$\mathcal{P}^\mathrm{M}_\mathrm{S}$ are isomorphic.

\textit{Proof.} 

(i) Let $(\alpha, \eta ) \in \mathcal{P}_\mathrm{M}$.
Since $\gamma(\eta) \in \bar{U}(2, \eta)$, we have according to
Theorem~7 that $(\alpha | \gamma(\eta) ) \in
\mathcal{P}_\mathrm{S}$.
This means that $\mathcal{P}^\mathrm{M}_\mathrm{S} \subset
\mathcal{P}_\mathrm{S}$.
The one-to-one feature of the mapping~(\ref{eq_sm2_mapp}) follows
from the one-to-one feature of the mapping $\eta \mapsto
\gamma(\eta)$, Eq.~(\ref{eq_sm2_gammadef}), which proves the first
part of the theorem.

(ii) First, the identity element $(1, 1) \in
\mathcal{P}_\mathrm{M}$  is mapped on the identity element
$(1 | 1 ) \in \mathcal{P}_\mathrm{S}$ which proves that the set
$\mathcal{P}^\mathrm{M}_\mathrm{S}$ contains the identity element.
Second, let $(\alpha_j | \beta_j ) \in
\mathcal{P}^\mathrm{M}_\mathrm{S}$ for $j = 1, 2$.
Then $\eta_j$ exists such, that $\beta_j = \gamma(\eta_j)$ and
$(\alpha_j , \eta_j ) \in \mathcal{P}_\mathrm{M}$ for $j = 1, 2$.
Let us define $(\alpha_3 | \beta_3 ) = (\alpha_1 | \beta_1 )
(\alpha_2 | \beta_2 ) \in \mathcal{P}_\mathrm{S}$ 
with the rule~(\ref{eq_sm2_gmrsp}), which yields
$\beta_3 = \beta_1 \beta_2$.
Then $(\alpha_1 , \eta_1 ) (\alpha_2 , \eta_2 ) =
(\alpha_3 , \eta_3 ) \in \mathcal{P}_\mathrm{M}$ with
$\eta_3 = \eta_1 \eta_2$ according to Eq.~(\ref{eq_sm2_gmrmp}).
Using the rule~(\ref{eq_sm2_gamma12}) we get $\beta_3 =
\gamma(\eta_3)$, which means that $(\alpha_3 | \beta_3 ) \in 
\mathcal{P}^\mathrm{M}_\mathrm{S}$, so that the product of two
elements of $\mathcal{P}^\mathrm{M}_\mathrm{S}$ also belongs to
$\mathcal{P}^\mathrm{M}_\mathrm{S}$.
Third, let $(\alpha | \beta ) \in
\mathcal{P}^\mathrm{M}_\mathrm{S}$, so that an $\eta$ exists such,
that $\beta = \gamma(\eta)$ and $(\alpha , \eta ) \in
\mathcal{P}_\mathrm{M}$.
Then the inverse element to $(\alpha , \eta )$ also belongs to
$\mathcal{P}_\mathrm{M}$, so that $(\alpha^{-1} , \eta )
\in \mathcal{P}_\mathrm{M}$, and, consequently, $(\alpha^{-1} |
\gamma(\eta) ) \in \mathcal{P}^\mathrm{M}_\mathrm{S}$.
However, we have $\gamma(\eta) = \gamma^{-1}(\eta) = \beta^{-1}$,
so that the element $(\alpha^{-1} | \beta^{-1} )$, which is the
inverse of the element $(\alpha | \beta )$, belongs to
$\mathcal{P}^\mathrm{M}_\mathrm{S}$ as well.
This completes the proof of the second part of the theorem.

(iii) Let $(\alpha_j , \eta_j ) \in \mathcal{P}_\mathrm{M}$ and
$(\alpha_j | \beta_j ) \in \mathcal{P}^\mathrm{M}_\mathrm{S}$,
where $\beta_j = \gamma(\eta_j)$ for $j = 1, 2, 3$.
Let $(\alpha_1 , \eta_1 ) (\alpha_2 , \eta_2 ) =
(\alpha_3 , \eta_3 )$ according to Eq.~(\ref{eq_sm2_gmrmp}).
Then $\eta_3 = \eta_1 \eta_2$ and, using the
rule~(\ref{eq_sm2_gamma12}),
$\beta_3 = \gamma(\eta_3) = \gamma(\eta_1) \gamma(\eta_2) =
\beta_1 \beta_2$, which means that $(\alpha_1 | \beta_1 )
(\alpha_2  | \beta_2 ) = (\alpha_3  | \beta_3 )$
according to Eq.~(\ref{eq_sm2_gmrsp}).
This proves the isomorphism and it completes the proof of the whole
theorem.

Let us formulate finally a theorem corresponding to the previous
Theorem~6.

\textbf{Theorem 9.}
Each element of the spin point group,
$(\alpha | \beta ) \in \mathcal{P}_\mathrm{S}$, can be written
as a product of two elements:
an element $( 1 | \beta_1 )$, where $\beta_1 \in \bar{U}(2, 1)$,
and an element of the group $\mathcal{P}^\mathrm{M}_\mathrm{S}$
isomorphic with the modified magnetic point group
$\mathcal{P}_\mathrm{M}$, $(\alpha | \beta_2 ) \in
\mathcal{P}^\mathrm{M}_\mathrm{S}$.

\textit{Proof.} 
Let $(\alpha | \beta ) \in \mathcal{P}_\mathrm{S}$, then (from
Theorem~7) an $\eta$ exists such that $\beta \in \bar{U}(2, \eta)$
and $(\alpha , \eta ) \in \mathcal{P}_\mathrm{M}$.
Let us define $\beta_2 = \gamma(\eta)$, so that $(\alpha | \beta_2 )
\in \mathcal{P}^\mathrm{M}_\mathrm{S}$ and $\beta_2 \in
\bar{U}(2, \eta)$.
Let us take further $\beta_1 = \beta \beta_2$ which yields  
$\beta_1 \in \bar{U}(2, 1)$.
Moreover, we get $\beta^2_2 = \gamma^2(\eta) = 1$ and 
$\beta_1 \beta_2 = \beta$, so that $( 1 | \beta_1 )
(\alpha | \beta_2 ) = (\alpha | \beta )$, which completes the proof.

\textit{Consequence.}
In analogy to Theorem~6, this theorem shows the relation among the
full spin point group $\mathcal{P}_\mathrm{S}$ and its two subgroups:
the group $\mathcal{P}^\mathrm{M}_\mathrm{S}$ (isomorphic with the
modified magnetic point group $\mathcal{P}_\mathrm{M}$) and the group
of elements of the type $( 1 | \beta )$, where $\beta$ is an element
of the system-independent group $\bar{U}(2, 1)$.

\subsection{Additional remarks}

The representation~(\ref{eq_sm2_biggamma}) of the big spin space
group $\mathcal{G}^\infty_\mathrm{S}$ and the
mapping~(\ref{eq_sm2_maps}) allow one to define a unitary
representation $\Delta( \alpha , \mathbf{t} , \eta )$ of the big 
modified magnetic space group $\mathcal{G}^\infty_\mathrm{M}$ as
\begin{equation}
\Delta( \alpha , \mathbf{t} ,  \eta ) = 
\Gamma( \alpha , \mathbf{t} | \gamma(\eta) ) .
\label{eq_sm2_bigdelta}
\end{equation}
Its action on the basic kets $| \mathbf{r}s \rangle$ comes out
simply as
\begin{equation}
\Delta( \alpha , \mathbf{t} ,  \eta ) | \mathbf{r}s \rangle =
| \alpha \mathbf{r} + \mathbf{t} , \eta s \rangle ,
\label{eq_sm2_deltars}
\end{equation}
which follows from $\gamma(\eta) | s \rangle = | \eta s \rangle$
valid for all values of $\eta$ ($\eta = \pm 1$) and $s$
($s = \pm 1$).
The transparent rule~(\ref{eq_sm2_deltars}) leads to the particular
representation of the magnetic point group $\mathcal{P}_\mathrm{M}$
introduced in Section II C of the main article.

Let us also discuss briefly the operators $\beta$ of the group
$\bar{U}(2, 1)$.
This continuous group has three generators: the unit operator $1$,
the spin operator $\sigma$, and the basic antiunitary operator
$\tau$ defined (for $s \in \{ 1 , -1 \}$) by
\begin{equation}
\tau | s \rangle = | s \rangle .
\label{eq_sm2_taudef}
\end {equation}
The properties of $\tau$ include relations
\begin{equation}
\tau^2 = 1 ,
\qquad
\sigma \tau = \tau \sigma ,
\qquad
\omega \tau = \tau \omega .
\label{eq_sm2_tauprop}
\end {equation}
All unitary operators $\beta \in \bar{U}(2, 1)$ can be parametrized
as $\beta'(p,q) = \exp[ i (p 1 + q \sigma) ]$, where the $p$ and $q$
are real numbers, while all antiunitary operators
$\beta \in \bar{U}(2, 1)$ can be written as 
$\beta''(p,q) = \beta'(p,q) \tau$.
The physical interpretation of all three generators is obvious: the
unit operator $1$ generates merely an arbitrary phase factor, the
spin operator $\sigma$ generates an arbitrary rotation in the spin
space around an axis parallel to the direction of magnetic moments
of the collinear magnet, and the antiunitary operator $\tau$
represents the time reversal of spin-zero particles moving in the
decoupled spin-up ($s = 1$) and spin-down ($s = -1$) channels.
The operator $\tau$ should carefully be distinguished from the
time-reversal operator of particles with spin 1/2;
the latter is given by $\sigma \omega \tau$ and it satisfies the
identity $( \sigma \omega \tau )^2 = -1$.
Note that the operator $\sigma \omega \tau$ does not belong to
$\bar{U}(2, 1)$, but $\sigma \omega \tau \in \bar{U}(2, -1)$.
This reflects the fact that the pure spin-1/2 time reversal
is not a symmetry element of the nonrelativistic collinear magnet,
$(1 , \mathbf{0} | \sigma \omega \tau ) \notin
\mathcal{G}_\mathrm{S}$, in contrast to the pure spin-zero
time reversal, $(1 , \mathbf{0} | \tau ) \in
\mathcal{G}_\mathrm{S}$. 
In view of Theorems~6 and 9, the operators $\sigma$ and $\tau$
should thus be considered (along with operators representing all
elements of the modified magnetic groups) in a complete
group-theoretical analysis of the studied quantum systems,
see Section II of the main article.


\providecommand{\noopsort}[1]{}\providecommand{\singleletter}[1]{#1}%

\end{document}